\documentclass[12pt,a4,sans]{article}
\pdfoutput=1
\usepackage[latin1]{inputenc}
\usepackage{amssymb}
\usepackage{amsmath}
\usepackage{amsthm}
\usepackage{amsfonts}
\usepackage[pdftex]{graphicx}
\usepackage{geometry}
\usepackage{braket}
\usepackage{enumerate}
\usepackage{subcaption}
\allowdisplaybreaks
\usepackage[table,usenames,dvipsnames]{xcolor}
\usepackage{setspace}
\usepackage[in]{fullpage}
\usepackage{hyperref}
\usepackage{wrapfig}
\usepackage[font=small,labelfont=bf]{caption}
\usepackage{mathtools}
\usepackage[nottoc,notlot,notlof]{tocbibind}
\usepackage[nosort]{cite}
\usepackage{parskip}
\usepackage{tikz}

\usetikzlibrary{calc,arrows, decorations.markings, shapes.misc, decorations.pathmorphing}

\def\boxsize{0.7}
\def\blobsize{0.55cm}
\def\bwblobsize{0.35cm}

\tikzset{
  ->-/.style={
    decoration={
      markings,
      mark=at position #1 with {\arrow{latex}}},
    postaction={decorate}
  },
  ->-/.default=0.5
}

\tikzset{
    wavy/.style={decorate, decoration={snake}, draw=red},
}

\tikzset{VO/.style={cross out, draw, 
         minimum size=5pt, 
         inner sep=0pt, outer sep=0pt}}

\tikzset{VOline/.style={decorate, decoration={snake}},
}

\tikzset{
    partial ellipse/.style args={#1:#2:#3}{
        insert path={+ (#1:#3) arc (#1:#2:#3)}
    }
}

\newcommand{\blackdot}{\node[circle, fill=black, draw, inner sep=0pt, minimum size=\bwblobsize]}
\newcommand{\whitedot}{\node[circle, fill=white, draw, inner sep=0pt, minimum size=\bwblobsize]}
\newcommand{\greyblob}{\node[circle, fill=black!20, draw, inner sep=2pt, minimum size=\blobsize]}

\newcommand{\drawULblack}{\blackdot (UL) at (-\boxsize,  \boxsize) {};}

\newcommand{\drawLRblack}{\blackdot (LR) at ( \boxsize, -\boxsize) {};}

\newcommand{\drawURwhite}{\whitedot (UR) at ( \boxsize,  \boxsize) {};}

\newcommand{\drawLLwhite}{\whitedot (LL) at (-\boxsize, -\boxsize) {};}

\newcommand{\drawUL}[1]{\greyblob (UL) at (-\boxsize,  \boxsize) {#1};}
\newcommand{\drawUR}[1]{\greyblob (UR) at ( \boxsize,  \boxsize) {#1};}
\newcommand{\drawLR}[1]{\greyblob (LR) at ( \boxsize, -\boxsize) {#1};}

\newcommand{\drawULempty}{\coordinate (UL) at (-\boxsize,  \boxsize);}
\newcommand{\drawURempty}{\coordinate (UR) at ( \boxsize,  \boxsize);}
\newcommand{\drawLRempty}{\coordinate (LR) at ( \boxsize, -\boxsize);}
\newcommand{\drawLLempty}{\coordinate (LL) at (-\boxsize, -\boxsize);}

\newcommand{\drawboxinternallines}{
  \draw (UL) -- (UR) -- (LR) -- (LL) -- (UL);
}

\newcommand{\drawregionvariables}[5]{
  \draw ( 0.0, -1.4) node {#1};
  \draw (-1.4,  0.0) node {#2};
  \draw ( 0.0,  1.4) node {#3};
  \draw ( 1.4,  0.0) node {#4};
  \draw ( 0.0,  0.0) node {#5};
}

\newcommand{\drawblobs}{
\node[circle, fill=black!20, draw, minimum size=0.6cm] (blobL) at (-1, 0) {};
\node[circle, fill=black!20, draw, minimum size=0.6cm] (blobR) at (+1, 0) {};
}

\newcommand{\drawupperlower}{
\draw (blobL) to[out=60, in=120] (blobR);
\draw (blobL) to[out=-60, in=-120] (blobR);
}

\newcommand{\drawupper}{
\draw (blobL) to[out=60, in=120] (blobR);
}

\newcommand{\drawlower}{
\draw (blobL) to[out=-60, in=-120] (blobR);
}

\newcommand{\drawuppercross}{
\draw (blobL) to[out=60, in=120] node[cross out, draw=red]{} (blobR);
}

\newcommand{\drawlowercross}{
\draw (blobL) to[out=-60, in=-120] node[cross out, draw=red]{} (blobR);
}

\newcommand{\drawuppercut}{
\draw (blobL) to[out=60, in=120] node[circle, fill=white, draw=white]{} (blobR);
\draw[red, dashed] (0, 0.5) -- (0, 0.9);
}

\newcommand{\drawlowercut}{
\draw (blobL) to[out=-60, in=-120] node[circle, fill=white, draw=white]{} (blobR);
\draw[red, dashed] (0,-0.5) -- (0,-0.9);
}

\newcommand{\drawlowerell}{
\draw (blobL) -- ++( -60:1) node[below=-2pt] {$-\ell$};
\draw (blobR) -- ++(-120:1) node[below left=-2pt] {$\ell$};
}

\newcommand{\drawupperell}{
\draw (blobL) -- ++(  60:1) node[above=-1pt] {$\ell$};
\draw (blobR) -- ++( 120:1) node[above=-2pt] {$-\ell\quad$};
}

\numberwithin{equation}{section}

\DeclareMathOperator*{\Res}{Res}

\hypersetup{
	colorlinks=true,
	linktoc=page,
	citecolor=Blue,
	linkcolor=Blue,
	urlcolor=Blue} 

\makeatletter
\renewcommand{\maketitle}
{ \begingroup \begin{center} \large {\bf \@title}
		\vskip 5pt \large \@author \\ \vskip 5pt \@date \end{center}
	\vskip 5pt \endgroup \setcounter{footnote}{0} }
\makeatother

\long\def\symbolfootnote[#1]#2{\begingroup%
	\def\thefootnote{\fnsymbol{footnote}}\footnote[#1]{#2}\endgroup}

\begin{document}

\begin{flushright}
	QMUL-PH-18-15\\
	HU-EP-18/25
\end{flushright}
	
\vspace{20pt}

\begin{center}		
	{\Large \bf Form factor recursion relations at loop level}
	
	\vspace{25pt}

	{\mbox {\bf  Lorenzo~Bianchi$^{a}$, Andreas~Brandhuber$^{a}$, 
		Rodolfo Panerai$^{a}$ and
		Gabriele~Travaglini$^{a,b\S}$}}
	\vspace{0.5cm}
		
	\begin{quote}
		{\small \em
		\begin{itemize}
		\item[\ \ \ \ \ \ $^a$]
		\begin{flushleft}
			Centre for Research in String Theory\\
			School of Physics and Astronomy\\
			Queen Mary University of London\\
			Mile End Road, London E1 4NS, United Kingdom
		\end{flushleft}			

	\item[\ \ \ \ \ \ $^b$]
		Institut f\"{u}r Physik und IRIS Adlershof\\
		Humboldt-Universit\"{a}t zu Berlin\\
		Zum Gro{\ss}en Windkanal 6, 12489 Berlin, Germany
		\end{itemize}
		}
	\end{quote}
		
	\vspace{15pt}
		
{\bf Abstract}
	\end{center}
	\vspace{0.3cm}
\noindent
\noindent
We introduce a prescription to define form factor integrands at loop level in planar $\mathcal{N}\!=\!4$ supersymmetric Yang-Mills theory.
This relies on  a periodic kinematic configuration that has been instrumental  to describe form factors at strong coupling in terms of periodic Wilson loops.  
With this prescription, we are able to formulate loop-level recursion relations for planar form factor integrands, using a two-line (BCFW) and an all-line shift. 
We also point out important differences with the known recursion relations of integrands of planar loop amplitudes.
We present a number of concrete one-loop examples to illustrate and validate our prescription for form factor integrands.
	\vfill
	\hrulefill
	\newline
\vspace{-1cm}
$^{\S}$~\!\!{\tt\footnotesize\{lorenzo.bianchi, a.brandhuber, r.panerai, g.travaglini\}@qmul.ac.uk}	
	
	\setcounter{page}{0}
	\thispagestyle{empty}
	\newpage
	
\setcounter{tocdepth}{4}
\hrule height 0.75pt
\tableofcontents
\vspace{0.8cm}
\hrule height 0.75pt
\vspace{1cm}
	
\setcounter{tocdepth}{2}

\section{Introduction}\label{sec:Introduction}

Over the past few years, great effort has been devoted  to extending powerful on-shell methods specifically developed for the computation of amplitudes to partially off-shell quantities such as form factors, and fully off-shell quantities such as correlation functions.
Form factors are the overlap of an $n$-particle state with a state produced by a local gauge-invariant operator $\mathcal{O}(x)$ applied on to the vacuum, and naturally appear as amplitudes in effective theories. Hence, one would expect that many of the amplitudes methods can be ported to this interesting case as well.
This expectation was confirmed in \cite{Brandhuber:2010ad, Brandhuber:2011tv} where unitarity cuts \cite{Bern:1994zx, Bern:1994cg}, BCFW recursion relations \cite{Britto:2004ap,Britto:2005fq}  and MHV diagrams \cite{Cachazo:2004kj,Brandhuber:2004yw} were used directly to find new expressions for tree-level and one-loop form factors. These papers also provided first indications of extensions of the amplitude/Wilson loop duality \cite{Alday:2007hr,Drummond:2007aua,Brandhuber:2007yx,Drummond:2007cf}
and formulations of form factors in momentum twistor space \cite{Hodges:2009hk}.%
\footnote{The corresponding periodic kinematic configurations (``periodic Wilson loops") in dual momentum space and in momentum twistor space play an important role in this paper.}

It also became clear soon after \cite{Brandhuber:2012vm} that more advanced methods like generalised unitarity \cite{Bern:2004cz,Britto:2004nc} and the symbol of transcendental functions \cite{Goncharov:2010jf} could be employed effectively to obtain a plethora of novel results \cite{Brandhuber:2014ica, Loebbert:2015ova, Brandhuber:2016fni, Loebbert:2016xkw, Brandhuber:2017bkg}.
It also turned out that new geometric formulations like Grassmannians \cite{ArkaniHamed:2009dn} and twistor strings could be extended, see \cite{Frassek:2015rka} and \cite{Koster:2016ebi, Koster:2016loo, Koster:2016fna}, respectively. Even the scattering equations \cite{Cachazo:2013hca}  and related formulations in twistor space \cite{Roiban:2004yf,Spradlin:2009qr} could be generalised to form factors \cite{1607.02843,1608.03277}. 
Explicit results for a number of helicity configurations and for super form factors of the stress-tensor multiplet operator were obtained, see  \cite{Brandhuber:2010ad, Brandhuber:2011tv,Bork:2014eqa, Bork:2016hst, Chicherin:2016qsf}, and in particular in \cite{Bork:2014eqa} an expression for the $n$-point NMHV form factors of this operator was obtained by solving the supersymmetric BCFW recursion relation \cite{ArkaniHamed:2008gz,Brandhuber:2008pf} in a similar way as done in \cite{Drummond:2008cr} for superamplitudes.

The successful extension of recursive techniques to integrands of planar loop amplitudes in $\mathcal{N}\!=\!4$ SYM was accomplished in \cite{ArkaniHamed:2010kv}, following earlier work of \cite{CaronHuot:2010zt}.
A key insight of \cite{ArkaniHamed:2010kv} is that at each loop order one can unambiguously define an object, the planar integrand, which can then be computed recursively. This relies on the fact that for a colour-ordered amplitude, one can re-write the momenta of the particles using region momenta as \cite{Drummond:2006rz,Drummond:2007aua}
\begin{align}
\label{regmom}
p_i = x_i - x_{i+1} \;. 
\end{align}
This change of variables automatically implements momentum conservation, and is a crucial ingredient in the duality between Wilson loops and scattering amplitudes in $\mathcal{N}\!=\!4$ SYM \cite{Alday:2007hr,Drummond:2007aua,Brandhuber:2007yx}. In this duality, the Wilson loop is stretched along a polygonal light-like contour which connects the points $x_i$. At strong coupling \cite{Alday:2007hr}, this mapping can be interpreted as a T-duality transformation on the $AdS_5$ coordinates. In the weak coupling picture \cite{Drummond:2007aua,Brandhuber:2007yx,Drummond:2007cf}, the assignment of region momenta for the planar integrand shows the emergence of an anomalous hidden symmetry, known as dual conformal invariance (DCI) \cite{Drummond:2007au,Drummond:2008vq}. In the Wilson loop picture, DCI is simply conformal invariance of the Wilson loop expectation value, which is anomalous due to the presence of cusps along the contour, with the anomaly being controlled by the cusp anomalous dimension. This interpretation allows to check DCI also on the integrated amplitude by applying dual conformal generators on the final result expressed in terms of region variables \cite{Drummond:2007au,Drummond:2008vq}.

For form factors, two important differences need to be taken into account. First, momentum conservation now reads
\begin{equation}
 \sum_{i=1}^n p_i = q \;,
\end{equation}
where $q$ is the incoming momentum of the off-shell leg associated with the operator insertion. This implies that the dual Wilson line cannot be drawn as a closed, piecewise light-like polygon. The proposal of \cite{Alday:2007he,Maldacena:2010kp} at strong coupling is to draw the dual contour as a periodic Wilson line, with period $q$.
Furthermore, the inserted local operator is gauge invariant, {\it i.e.}\ a colour singlet, thus making the object inherently non-planar. In \cite{Brandhuber:2010ad}, a similar picture was advocated  at weak coupling, and further discussed in Section 5 of \cite{Brandhuber:2011tv}. In the latter paper, dual MHV rules which crucially involve a periodic configuration in momentum twistor space were also introduced, and applied to the computation of tree-level and one-loop supersymmetric form factors of protected operators.

In this work we leave aside a more detailed definition of the form factor/Wilson line duality, and instead give a well-motivated prescription for expressing  form factors in terms of region variables living on a periodic contour. Crucially, with such a prescription one can unambiguously define  one-loop integrands even for form factors, and hence study loop recursion relations, both for a two-line and an all-line shift.
With this prescription in hand, recursion relations can be formulated straightforwardly.
Furthermore, and importantly, this prescription is mandatory in order to define and understand the action of  dual conformal symmetry on form factors. In the present paper we will define this prescription and use it to study loop-level recursion relations for form factor integrands, while the realisation of DCI will be fully studied in the companion paper \cite{Part2}. Throughout these works we will consider form factors of the (chiral part of the) stress-tensor multiplet operator. It would clearly be of interest to extend our discussion to more general operators as well.

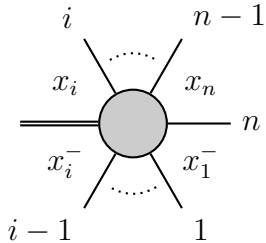
\begin{figure}[htbp]
\centering
\begin{tikzpicture}[thick]
  \node[circle, fill=black!20, draw, minimum size=0.9cm] (blob) at (0.0,  0.0) {};
  \draw[double] (blob) -- ++( 180:1.5) node[anchor=east] {};
  \draw (blob) -- ++( 120:1.3) node[anchor=south east] {$i$};
  \draw[dotted] (blob)+( 110:0.9) to [bend left=30] ++(  70:0.9);
  \draw (blob) -- ++(  60:1.3) node[anchor=south west] {$n-1$};
  \draw (blob) -- ++(   0:1.3) node[anchor=west] {$n$};
  \draw (blob) -- ++( -60:1.3) node[anchor=north west] {$1$};
  \draw[dotted] (blob)+( -70:0.9) to [bend left=30] ++(-110:0.9);
  \draw (blob) -- ++(-120:1.3) node[anchor=north east] {$i-1$};
  \node[] (xi) at (-0.9, 0.5) {$x_i$};
  \node[] (xn) at ( 0.9, 0.5) {$x_n$};
  \node[] (x1m) at ( 0.9,-0.5) {$x_1^-$};
  \node[] (xim) at (-0.9,-0.5) {$x_i^-$};
\end{tikzpicture}
 \caption{Possible assignments of region momenta in a planar form factor diagram. The double line corresponds to the off-shell leg carrying incoming momentum $q$. In our notation $x_i^-\equiv x_i-q$.}\label{arbdiag}
\end{figure}
The rest of the paper is organised as follows.  In Section \ref{regionvar} we discuss the assignment of region momenta for form factors and introduce a periodic kinematic configuration inspired by \cite{Alday:2007he,Maldacena:2010kp,Brandhuber:2010ad}. This is a key step which then allows us to formulate recursion relations.  In Section \ref{review} we review NMHV form factors and the particular $R$-invariants used to express them, some of which are novel compared to amplitudes. Section \ref{recrelloop} is the central section of the paper. There we introduce two types of recursion relations, 
namely the two-line shift, or BCFW recursion relation for the loop integrand, and the all-line shift recursion relation, which is
equivalent to MHV diagrams. Several  one-loop examples are described in order to illustrate the practical implementation of the recursions and point out important differences compared to recursion relations for amplitudes. Finally, in two appendices we describe our conventions and present details of the derivation of NMHV tree-level form factors.

\section{Assignment of region momenta for form factors}\label{regionvar}
We begin our discussion by considering a generic form factor diagram, such as that in Figure~\ref{arbdiag}, contributing in the planar limit. This could be a Feynman or BCFW diagram or an integral function, and we colour order all external on-shell legs. Because the operator is a gauge singlet, the corresponding line $q$ can be inserted between any pair of lines. Up to one loop one can only have planar diagrams, but starting from two loops, non-planar integrals can appear even at leading order  in colour. 

Once we have drawn $q$ in a particular position, {\it e.g.}~between legs $i-1$ and $i$ as in Figure~\ref{arbdiag}, we label  the region variables starting from $q$ and moving in a clockwise fashion. We then introduce the region momenta as in \eqref{regmom}, with  the identification 
\begin{align}
x_{i+n} = x_i - q \equiv x_i^- \;. 
\end{align}
When we get back to the leg with momentum $q$, we have moved all the way to $x_i^-$ and this provides a natural way to rewrite $q$ in terms of region variables as%
\footnote{In our conventions, the momentum $q$ is incoming.}
$q=x_i-x_{i}^-$.  
 
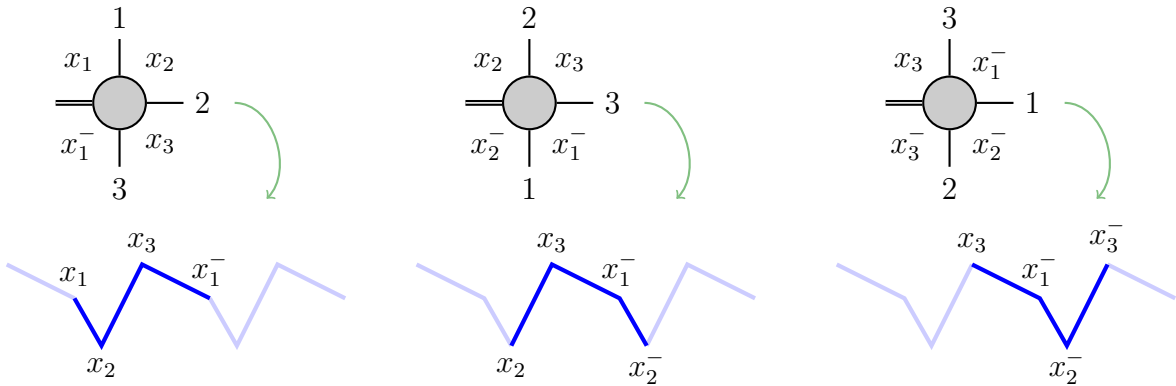
\begin{figure}[htbp]
\begin{subfigure}[t]{.33\linewidth}\vspace{0pt}
\centering
\begin{tikzpicture}[thick]
\begin{scope}[xshift=0.6cm ,yshift=2.6cm, scale=0.85]
\node[circle, fill=black!20, draw, minimum size=0.7cm] (blob) at (0.0,  0.0) {};
\draw[double] (blob) -- ++( 180:1.0) node[left] {};
\draw (blob) -- ++( 90:1.0) node[above] {$1$};
\draw (blob) -- ++(  0:1.0) node[right] {$2$};
\draw (blob) -- ++(-90:1.0) node[below] {$3$};
\node[] at ( 135:0.9) {$x_1$};
\node[] at (  45:0.9) {$x_2$};
\node[] at ( -45:0.9) {$x_3$};
\node[] at (-135:0.9) {$x_1^-$};
\draw[green!50!black!50,->] (1.8,0) to[out=0, in=45] (2.3,-1.5);
\end{scope}
\begin{scope}[scale=0.9]
\draw[ultra thick, blue!20] (-1,0.5) -- (0,0);
\draw[ultra thick, blue!20] (2,0) -- (2.4,-0.7) -- (3,0.5) -- (4,0);
\draw[ultra thick, blue] (0,0) node[above, black] {$x_1$} -- (0.4,-0.7) node[below, black] {$x_2$} -- (1,0.5) node[above, black] {$x_3$} -- (2,0) node[above, black] {$x_1^-$};
\end{scope}
\end{tikzpicture}
\end{subfigure}%
\begin{subfigure}[t]{.33\linewidth}\vspace{0pt}
\centering
\begin{tikzpicture}[thick]
\begin{scope}[xshift=0.6cm ,yshift=2.6cm, scale=0.85]
\node[circle, fill=black!20, draw, minimum size=0.7cm] (blob) at (0.0,  0.0) {};
\draw[double] (blob) -- ++( 180:1.0) node[left] {};
\draw (blob) -- ++( 90:1.0) node[above] {$2$};
\draw (blob) -- ++(  0:1.0) node[right] {$3$};
\draw (blob) -- ++(-90:1.0) node[below] {$1$};
\node[] at ( 135:0.9) {$x_2$};
\node[] at (  45:0.9) {$x_3$};
\node[] at ( -45:0.9) {$x_1^-$};
\node[] at (-135:0.9) {$x_2^-$};
\draw[green!50!black!50,->] (1.8,0) to[out=0, in=45] (2.3,-1.5);
\end{scope}
\begin{scope}[scale=0.9]
\draw[ultra thick, blue!20] (-1,0.5) -- (0,0) -- (0.4,-0.7);
\draw[ultra thick, blue!20] (2.4,-0.7) -- (3,0.5) -- (4,0);
\draw[ultra thick, blue] (0.4,-0.7) node[below, black] {$x_2$} -- (1,0.5) node[above, black] {$x_3$} -- (2,0) node[above, black] {$x_1^-$} -- (2.4,-0.7) node[below=-3pt, black] {$x_2^-$};
\end{scope}
\end{tikzpicture}
\end{subfigure}
\begin{subfigure}[t]{.33\linewidth}\vspace{0pt}
\centering
\begin{tikzpicture}[thick]
\begin{scope}[xshift=0.6cm ,yshift=2.6cm, scale=0.85]
\node[circle, fill=black!20, draw, minimum size=0.7cm] (blob) at (0.0,  0.0) {};
\draw[double] (blob) -- ++( 180:1.0) node[left] {};
\draw (blob) -- ++( 90:1.0) node[above] {$3$};
\draw (blob) -- ++(  0:1.0) node[right] {$1$};
\draw (blob) -- ++(-90:1.0) node[below] {$2$};
\node[] at ( 135:0.9) {$x_3$};
\node[] at (  45:0.9) {$x_1^-$};
\node[] at ( -45:0.9) {$x_2^-$};
\node[] at (-135:0.9) {$x_3^-$};
\draw[green!50!black!50,->] (1.8,0) to[out=0, in=45] (2.3,-1.5);
\end{scope}
\begin{scope}[scale=0.9]
\draw[ultra thick, blue!20] (-1,0.5) -- (0,0) -- (0.4,-0.7) -- (1,0.5);
\draw[ultra thick, blue!20] (3,0.5) -- (4,0);
\draw[ultra thick, blue] (1,0.5) node[above, black] {$x_3$} -- (2,0) node[above, black] {$x_1^-$} -- (2.4,-0.7) node[below=-3pt, black] {$x_2^-$} -- (3,0.5) node[above, black] {$x_3^-$};
\end{scope}
\end{tikzpicture}
\end{subfigure}
\caption{Form factor with three external legs and periodic dual configuration. The highlighted region is the one we select.}\label{example3leg}
\end{figure}

We would like to stress that the peculiarity of our prescription is that the definition of $q$ in terms of region variables changes according to the diagram we are considering, since a priori the off-shell leg is not ordered with respect to the on-shell ones. In the companion paper \cite{Part2}  we will show how this assignment is crucial in defining the action of dual conformal symmetry. 
In other words, given the infinite sequence of light-like segments in the periodic dual configuration, we associate to every diagram a particular period therein. As an example, in Figure~\ref{example3leg} we consider the three-leg case and show how the three possible configurations are mapped to three different periods.

Notice that our prescription involves the choice of an origin. For instance, in the first diagram of Figure~\ref{example3leg} we chose to start labelling regions from $x_1$ and then move clockwise around the diagram.
It should be clear that we could have labelled region momenta starting from any other vertex. This would have no consequences for the integrated result thanks to translation invariance in dual space. 
Nevertheless this choice has consequences in the definition of the loop integrand, and the action of the dual conformal generators.

The application of recursion relations to the loop integrand of scattering amplitudes requires  the unambiguous definition of the integrand itself. This is obtained in the planar limit by introducing region variables. In a similar way we can introduce region variables for the form factor loop integrands. At one loop they will involve propagators of the type $1/x_{0i}^2$, where $x_0$ is the region of the loop momentum. It  is also clear that an overall shift of the external region variables $x_i\to x_i+m q$ can be compensated by a shift in the loop variable $x_0\to x_0+mq$. This feature will be crucial in the derivation of the loop recursion relation presented later.
 
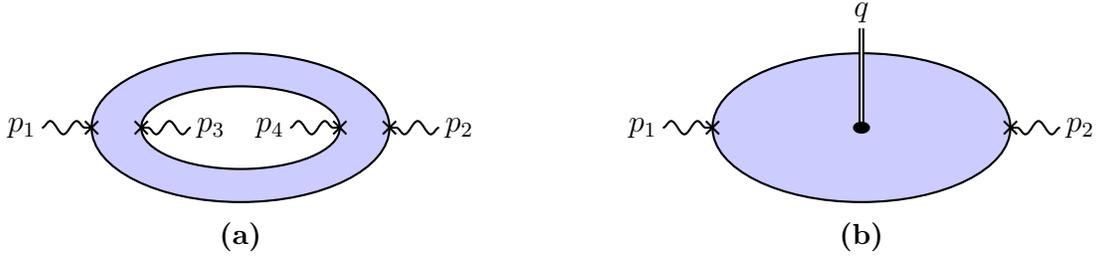
\begin{figure}[htbp]
\begin{subfigure}[t]{.5\linewidth}
\centering
\begin{tikzpicture}[thick, scale=1.1]
\draw[fill=blue!20] (0,0) ellipse (1.8cm and 0.9cm);
\draw[fill=white] (0,0) ellipse (1.2cm and 0.5cm);
\draw[VOline] (-2.4,0) node[left=-2pt] {$p_1$} to (-1.8,0) node[VO] {};
\draw[VOline] (-0.6,0) node[right=-2pt] {$p_3$} to (-1.2,0) node[VO] {};
\draw[VOline] (+0.6,0) node[left=-2pt] {$p_4$} to (+1.2,0) node[VO] {};
\draw[VOline] (+2.4,0) node[right=-2pt] {$p_2$} to (+1.8,0) node[VO] {};
\end{tikzpicture}
\caption{}
\end{subfigure}%
\begin{subfigure}[t]{.5\linewidth}
\centering
\begin{tikzpicture}[thick, scale=1.1]
\draw[fill=blue!20] (0,0) ellipse (1.8cm and 0.9cm);
\draw[VOline] (-2.4,0) node[left=-2pt] {$p_1$} to (-1.8,0) node[VO] {};
\draw[VOline] (+2.4,0) node[right=-2pt] {$p_2$} to (+1.8,0) node[VO] {};
\draw[double] (0,0) to (0,1.2) node[above=-2pt] {$q$};
\draw[fill] (0,0) ellipse (0.09cm and 0.06cm);
\end{tikzpicture}
\caption{}
\end{subfigure}
\caption{$(a)$ Worldsheet configuration for a four-point double-trace amplitude. Each ${\times}$ stands for the insertion of an open string vertex operator. $(b)$ Worldsheet configuration for the Sudakov form factor. The $\bullet$ stands for the closed string vertex operator.} \label{strings}
\end{figure}

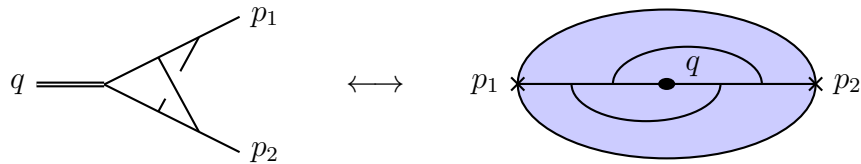
\begin{figure}[htbp]
\[
\raisebox{-.46\height}{
\begin{tikzpicture}[thick, scale=0.9]
\draw[double] (0,0) -- (-1,0) node[left] {$q$};
\draw (0,0) -- (2,1) node[right] {$p_1$};
\draw (0,0) -- (2,-1) node[right] {$p_2$};
\draw (0.8,-0.4) -- (1.4,0.7);
\draw (1,0) node[circle, fill=white, draw=white]{};
\draw (1.4,-0.7) -- (0.8,0.4);
\end{tikzpicture}
}
\hspace{0.5cm}\longleftrightarrow\hspace{0.5cm}
\raisebox{-.45\height}{
\begin{tikzpicture}[thick, scale=1.1]
\draw[fill=blue!20] (0,0) ellipse (1.8cm and 0.9cm);
\node[VO, label={180:$p_1$}] at (-1.8,0) {};
\node[VO, label={  0:$p_2$}] at (+1.8,0) {};
\draw[fill] (0,0) ellipse (0.09cm and 0.06cm);
\node[label={ 3:$q$}] at (-.05,-.05) {};
\draw (-1.8,0) -- (+1.8,0);
\draw (+0.25,0) [partial ellipse=0:180:0.9cm and 0.45cm];
\draw (-0.25,0) [partial ellipse=0:-180:0.9cm and 0.45cm];
\end{tikzpicture}
}
\]
\caption{A non-planar Feynman diagram which appears as planar when drawn on a punctured disk. All such diagrams contribute to the large $N$ form factor.}\label{non-planar}
\end{figure}

This property of the loop integrand can also be viewed   in the light of the recent work \cite{Ben-Israel:2018ckc}, where a Wilson loop dual for double-trace contributions to scattering amplitudes is discussed. Their main observation is based on the idea that the string worldsheet for double-trace  amplitudes has the topology of a cylinder, or equivalently of an annulus with open string insertions on the two boundaries. The case of form factors can be thought of as a degenerate limit of the double-trace amplitude, where the internal circle of the annulus shrinks to a point corresponding to a closed string insertion (see Figure \ref{strings}). 
In the large-$N$ limit on the gauge theory side only diagrams survive that can be drawn on the punctured disk topology.
A neat example is shown in Figure \ref{non-planar}, where a two-loop ``non-planar'' diagram contributing to the Sudakov form factor is drawn as a planar diagram on the punctured disk.%
\footnote{The degree of non-planarity of form factors is similar to the one described in \cite{Bern:2018oao}, since they can be made planar by removing the leg carrying momentum $q$. However $q$ is not light-like, hence the argument of \cite{Bern:2018oao} does not apply here. Actually we will show in the companion paper  \cite{Part2}  that the full dual conformal symmetry is preserved by form factor diagrams.}

The authors of \cite{Ben-Israel:2018ckc} established a correspondence between a double periodic Wilson loop and what they called the \emph{cylinder cut} of the amplitude. We refer to \cite{Ben-Israel:2018ckc} for the precise definition of the cylinder cut. Here we only point out that it depends on an additional momentum $\ell$, which, in the Wilson line picture, parameterises the distance between the two periodic Wilson lines. This momentum $\ell$, as much as our $x_0$ loop variable, is characterised by an ambiguity under shifts by an integer number of periods, {\it i.e.}~$\ell \mapsto \ell+mq$. In that case, the authors decided to eliminate this ambiguity by summing over all possible shifts.
For our purposes, instead of resolving the residual ambiguity of the integrand by performing an analogous sum, we just rely on the obvious property
\begin{equation}
\int \mathrm{d}^4x_0 \; f(x_0) = \int \mathrm{d}^4x_0 \; f(x_0+mq) \;,
\end{equation}
and regard different representations of the integrand related by shifts in $x_0$ as different representatives of the same equivalence class of integrands.
Although this introduces a level of freedom in defining integrand representations,  it allows to re-express the result of the recursion in terms of  a more conventional basis of integral functions.

\section{Overview of NMHV form factors}\label{review}

The modern approach to the study of scattering amplitudes is based on the idea that on-shell quantities can be used as building blocks for the construction of tree-level amplitudes as well as loop-level integrands. With the aim of applying a similar philosophy to the case of form factors, we review some existing results for planar NMHV form factors at tree and one-loop level. We start by setting our conventions. We denote by $F^{(l)}_{n,k}$ the $n$-point, $l$-loop $\mathrm{N}^k\mathrm{MHV}$ form factor and, analogously, with $A^{(l)}_{n,k}$ the $n$-point, $l$-loop $\mathrm{N}^k\mathrm{MHV}$ amplitude. At tree level, we graphically represent these quantities as
\begin{align}
F^{(0)}_{n,k} &=
\raisebox{-.45\height}{
\begin{tikzpicture}[thick]
  \greyblob (blob) at (0,0) {$\scriptstyle{k}$};
  \draw[double] (blob) -- ++( 180:0.8) node[left] {};
  \draw (blob) -- ++( 135:0.8) node[above left=-2pt] {$1$};
  \draw (blob) -- ++(  45:0.8) node[above right=-2pt] {$2$};
  \draw[dotted] (blob)+(  35:0.6) to [bend left=45] ++( -35:0.6);
  \draw (blob) -- ++( -45:0.8) node[below right=-4pt] {$n-1$};
  \draw (blob) -- ++(-135:0.8) node[below left=-2pt] {$n$};
\end{tikzpicture}
} &
A^{(0)}_{n,k} &=
\raisebox{-.45\height}{
\begin{tikzpicture}[thick]
  \greyblob (blob) at (0,0) {$\scriptstyle{k}$};
  \draw (blob) -- ++( 135:0.8) node[above left=-2pt] {$1$};
  \draw (blob) -- ++(  45:0.8) node[above right=-2pt] {$2$};
  \draw[dotted] (blob)+(  35:0.6) to [bend left=45] ++( -35:0.6);
  \draw (blob) -- ++( -45:0.8) node[below right=-4pt] {$n-1$};
  \draw (blob) -- ++(-135:0.8) node[below left=-2pt] {$n$};
\end{tikzpicture}
}
\end{align}
Notice that the number $k$ inside a circle indicates the $\mathrm{MHV}$ degree and we associate the label $k\!=\! -1$ with the three-point $\overline{\mathrm{MHV}}$ amplitude. We also use the following conventions for the particular cases of three-point tree-level amplitudes, and two-point tree-level form factor,
\begin{align}
A_{3,0}^{(0)} &=
\raisebox{-.44\height}{
\begin{tikzpicture}[thick]
  \blackdot (MHVb) at (0.0, 0.0) {};
  \draw (MHVb) -- ++( 120:0.8) node[anchor=east] {$1$};
  \draw (MHVb) -- ++(   0:0.8) node[anchor=west] {$2$};
  \draw (MHVb) -- ++(-120:0.8) node[anchor=east] {$3$};
\end{tikzpicture}
} &
A_{3,-1}^{(0)} &=
\raisebox{-.44\height}{
\begin{tikzpicture}[thick]
  \whitedot (MHVb) at (0.0, 0.0) {};
  \draw (MHVb) -- ++( 120:0.8) node[anchor=east] {$1$};
  \draw (MHVb) -- ++(   0:0.8) node[anchor=west] {$2$};
  \draw (MHVb) -- ++(-120:0.8) node[anchor=east] {$3$};
\end{tikzpicture}
} &
F_{2,0}^{(0)} &=
\raisebox{-.44\height}{
\begin{tikzpicture}[thick]
  \coordinate (FF) at (0.0, 0.0);
  \draw (FF) -- ++(  60:0.8) node[anchor=west] {$1$};
  \draw (FF) -- ++( -60:0.8) node[anchor=west] {$2$};
  \draw[double] (FF) -- ++( 180:0.8) node[anchor=east] {};
\end{tikzpicture}
} 
\end{align}
Explicit expressions for these quantities are given in Appendix~\ref{conventions}. These are the building blocks for the construction of the so-called $R$-invariants. The latter were defined, for the case of scattering amplitudes, as the dual conformal invariant quantities entering the ratio $A^{(0)}_{n,1}/A^{(0)}_{n,0}$ \cite{Drummond:2008vq,Brandhuber:2008pf}.  Subsequently, it became clear that the $R$-invariants determine the amplitude for any helicity configuration \cite{Britto:2005fq,Drummond:2008cr}. They  can be related to maximal cuts of one-loop $n$-point amplitudes using the BCFW bridge \cite{Britto:2004nc,Britto:2005fq} and recursive arguments \cite{Drummond:2008bq}, which can be better understood in twistor variables \cite{Mason:2009qx} or in the on-shell diagram formulation \cite{ArkaniHamed:2012nw}.  

\begin{figure}
\begin{align}
\raisebox{-.49\height}{
\begin{tikzpicture}[thick, scale=0.8]
  \drawLLwhite
  \drawUL{$\scriptstyle{0}$}
  \drawUR{$\scriptstyle{0}$}
  \drawLR{$\scriptstyle{0}$}
  \drawboxinternallines
  \drawregionvariables{$x_c,\theta_c$}{$\hspace{-1cm}x_{c+1},\theta_{c+1}$}{$x_a,\theta_a$}{\hspace{.2cm}$x_b,\theta_b$}{}
  \draw (UL) -- ++( 150:1.0) node[anchor=south east] {$\{P,\mathrm{q}_P\}$};
  \draw[dotted] (UL)+( 145:0.8) to [bend left=30] ++( 125:0.8);
  \draw (UL) -- ++( 120:1.0);
  \draw (UR) -- ++(  30:1.0) node[anchor=south west] {$\{Q,\mathrm{q}_Q\}$};
  \draw[dotted] (UR)+(  55:0.8) to [bend left=30] ++(  35:0.8);
  \draw (UR) -- ++(  60:1.0);
  \draw (LR) -- ++( -30:1.0) node[anchor=north west] {$\{R,\mathrm{q}_R\}$};
  \draw[dotted] (LR)+( -35:0.8) to [bend left=30] ++( -55:0.8);
  \draw (LR) -- ++( -60:1.0);
  \draw (LL) -- ++(-135:1.0);
  \draw[white] (LL) -- ++(-150:1.0) node[anchor=north east] {$\textcolor{black}{\{r,\mathrm{q}_r\}}$};
\end{tikzpicture}
} \; \nonumber
\end{align}
\caption{Conventions for assigning outgoing momenta and supermomenta as well as region variables for a generic kinematic configuration.}
\label{regions} 
\end{figure}
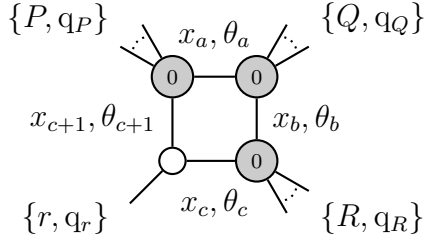

The extension to form factors was discussed in \cite{Bork:2014eqa}, where it was shown that the NMHV form factor can be expressed in terms of two types of $R$-invariants, which we introduce with the following on-shell diagrams:
\begin{align}\label{boxfunctions}
R'_{rst} =
\raisebox{-.48\height}{
\begin{tikzpicture}[thick, scale=0.8]
  \drawLLwhite
  \drawUL{$\scriptstyle{0}$}
  \drawUR{$\scriptstyle{0}$}
  \drawLR{$\scriptstyle{0}$}
  \drawboxinternallines
  \draw (UL) -- ++( 180:0.8) node[anchor=east] {$r+1$};
  \draw[dotted] (UL)+( 170:0.6) to [bend left=45] ++( 100:0.6);
  \draw (UL) -- ++(  90:0.8) node[anchor=south] {$s-1$};
  \draw[double] (UR) -- ++(   0:0.8) node[] {};
  \draw (UR) -- ++(  45:0.8) node[anchor=south west] {$t-1$};
  \draw[dotted] (UR)+(  80:0.6) to [bend left=45] ++(  50:0.6);
  \draw (UR) -- ++(  90:0.8) node[anchor=south] {$s$};
  \draw (LR) -- ++(   0:0.8) node[anchor=west] {$t$};
  \draw[dotted] (LR)+( -10:0.6) to [bend left=45] ++( -80:0.6);
  \draw (LR) -- ++( -90:0.8) node[anchor=north] {$r-1$}; 
  \draw (LL) -- ++(-135:0.8) node[anchor=north east] {$r$};
\end{tikzpicture}
} \;, \qquad
R''_{rst} =
\raisebox{-.48\height}{
\begin{tikzpicture}[thick, scale=0.8]
  \drawLLwhite
  \drawUL{$\scriptstyle{0}$}
  \drawUR{$\scriptstyle{0}$}
  \drawLR{$\scriptstyle{0}$}
  \drawboxinternallines
  \draw (UL) -- ++( 180:0.8) node[anchor=east] {$r+1$};
  \draw[dotted] (UL)+( 170:0.6) to [bend left=45] ++( 100:0.6);
  \draw (UL) -- ++(  90:0.8) node[anchor=south] {$s-1$};
  \draw (UR) -- ++(   0:0.8) node[anchor=west] {$t-1$};
  \draw[dotted] (UR)+(  80:0.6) to [bend left=45] ++(  10:0.6);
  \draw (UR) -- ++(  90:0.8) node[anchor=south] {$s$};
  \draw (LR) -- ++(   0:0.8) node[anchor=west] {$t$};
  \draw[dotted] (LR)+( -10:0.6) to [bend left=45] ++( -40:0.6);
  \draw (LR) -- ++( -45:0.8) node[anchor=north west] {$r-1$};
  \draw[double] (LR) -- ++( -90:0.8) node[] {};
  \draw (LL) -- ++(-135:0.8) node[anchor=north east] {$r$};
\end{tikzpicture}
} \;.
\end{align}
The precise relation between the above on-shell diagrams and the associated maximal cuts $\mathcal{C}'_{rst}$ and $\mathcal{C}''_{rst}$ reads
\begin{align}
\mathcal{C}^{\bullet}_{rst} = \mathrm{i} \Delta \, \frac{\delta^{(8)}(\mathrm{q})}{\langle 1\,2 \rangle \langle 2\,3 \rangle \cdots \langle n\,1 \rangle} \, R^{\bullet}_{rst} \;,
\end{align}
where the $\bullet$ indicates that this formula applies to both types of $R$-invariants, and 
\begin{align}\label{Delta}
\Delta = (p_r+P)^2(p_r+R)^2 - P^2R^2 \;.
\end{align}
Furthermore, we denote the total outgoing momentum and supermomentum in the upper-left, upper-right and lower-right corners respectively as $\{P,\mathrm{q}_P\}$, $\{Q,\mathrm{q}_Q\}$ and $\{R,\mathrm{q}_R\}$ (see Figure \ref{regions}).

A simple computation allows to derive an explicit expression for $R^{\bullet}_{rst}$, which can be straightforwardly applied to the case of form factors if $s \neq t$ \cite{Bork:2014eqa,Bork:2012tt},
\begin{align}\label{Rrst}
R^{\bullet}_{rst} = \frac{\langle s-1\,s\rangle \langle t-1\,t\rangle \, \delta^{(4)}(\langle \mathrm{q}_r+\mathrm{q}_P|QR|r\rangle  - \langle \mathrm{q}_R|QP|r\rangle )}{Q^2\langle r|RQ|s-1\rangle \langle r|RQ|s\rangle \langle r|PQ|t-1\rangle \langle r|PQ|t\rangle } \;.
\end{align}
In particular, no modification is needed for the corner case
\begin{align}
R''_{rsr} =
\raisebox{-.48\height}{
\begin{tikzpicture}[thick, scale=0.8]
  \drawLLwhite
  \drawUL{$\scriptstyle{0}$}
  \drawUR{$\scriptstyle{0}$}
  \drawLRempty
  \drawboxinternallines
  \draw (UL) -- ++( 180:0.8) node[anchor=east] {$r+1$};
  \draw[dotted] (UL)+( 170:0.6) to [bend left=45] ++( 100:0.6);
  \draw (UL) -- ++(  90:0.8) node[anchor=south] {$s-1$};
  \draw (UR) -- ++(   0:0.8) node[anchor=west] {$r-1$};
  \draw[dotted] (UR)+(  80:0.6) to [bend left=45] ++(  10:0.6);
  \draw (UR) -- ++(  90:0.8) node[anchor=south] {$s$};
  \draw[double] (LR) -- ++( -45:0.8);
  \draw (LL) -- ++(-135:0.8) node[anchor=north east] {$r$};
\end{tikzpicture}
} \;,
\end{align}
which does not have a counterpart in the context of amplitudes. However, the previous formula does not apply to the specific case $s=t$:
\begin{align}
R'_{rss} =
\raisebox{-.48\height}{
\begin{tikzpicture}[thick, scale=0.8]
  \drawLLwhite
  \drawUL{$\scriptstyle{0}$}
  \drawURempty
  \drawLR{$\scriptstyle{0}$}
  \drawboxinternallines
  \draw (UL) -- ++( 180:0.8) node[anchor=east] {$r+1$};
  \draw[dotted] (UL)+( 170:0.6) to [bend left=45] ++( 100:0.6);
  \draw (UL) -- ++(  90:0.8) node[anchor=south] {$s-1$};
  \draw[double] (UR) -- ++(  45:0.8) node {};
  \draw (LR) -- ++(   0:0.8) node[anchor=west] {$s$};
  \draw[dotted] (LR)+( -10:0.6) to [bend left=45] ++( -80:0.6);
  \draw (LR) -- ++( -90:0.8) node[anchor=north] {$r-1$}; 
  \draw (LL) -- ++(-135:0.8) node[anchor=north east] {$r$};
\end{tikzpicture}
} \;,
\end{align}
for which the correct result turns out to be given by
\begin{align}\label{Rrss}
R'_{rss} = -\frac{\langle s-1\,s\rangle \, \delta^{(4)}(\langle \mathrm{q}_r+\mathrm{q}_P|QR|r\rangle  - \langle \mathrm{q}_R|QP|r\rangle )}{Q^4\langle r|RQ|s-1\rangle \langle r|PQ|s\rangle \langle r|PQ|r\rangle} \;. 
\end{align}
The box diagrams are decorated with the assignment of specific region variables according to the rule outlined in Section \ref{regionvar}: we proceed clockwise and assign the $x$ and $\theta$ variables associated to each one of the four regions starting from the one that comes after the corner where the off-shell leg is inserted. We can represent this for a generic box diagram, as shown in Figure \ref{regions}, without the need to specify where the off-shell leg sits. By comparison with the diagrams in \eqref{boxfunctions} we have that
\begin{align}
x_c &\sim x_r \;, & \theta_c &\sim \theta_r \;, \cr
x_a &\sim x_s \;, & \theta_a &\sim \theta_s \;, \cr
x_b &\sim x_t \;, & \theta_b &\sim \theta_t \;,
\end{align}
where the $\sim$ sign indicates that the identity holds up to an appropriate shift by some integer multiple of a period. It is important to rewrite the results introduced so far in terms of region variables for the purpose of establishing recursion relations at loop level discussed in this paper, and to associate to each diagram a well defined behaviour under dual conformal transformations described in detail in the companion paper  \cite{Part2}.
In terms of region variables, one can rewrite \eqref{Rrst} and \eqref{Rrss} as
\begin{align}
R^{\bullet}_{rst} &= \frac{\langle s-1\,s\rangle \langle t-1\,t\rangle \, \delta^{(4)}(\langle r|x_{ca}x_{ab}|\theta_{br}\rangle  + \langle r|x_{cb}x_{ba}|\theta_{ar}\rangle )}{x_{ab}^2 \langle r|x_{cb}x_{ba}|s-1\rangle  \langle r|x_{cb}x_{ba}|s\rangle  \langle r|x_{ca}x_{ab}|t-1\rangle  \langle r|x_{ca}x_{ab}|t\rangle } \;, \label{Rfunreggen}\\
R'_{rss} &= -\frac{\langle s-1\,s\rangle \, \delta^{(4)}(\langle r|x_{ca}x_{ab}|\theta_{br}\rangle  + \langle r|x_{cb}x_{ba}|\theta_{ar}\rangle )}{x_{ab}^4 \langle r|x_{cb}x_{ba}|s-1\rangle  \langle r|x_{ca}x_{ab}|s\rangle  \langle r|x_{ca}x_{bc}|r\rangle } \; .\label{Rfunregspec}
\end{align}

Finally, we wish to present the complete tree-level, $n$-point NMHV form factor.
In \cite{Bork:2012tt} it was shown that the tree-level NMHV form factor can be written as a combination of $R$-invariants. In Appendix \ref{detailsNMHV} we give details of this derivation. The general idea is that, for $n$ particles, the BCFW recursion relation contains $2n-5$ diagrams involving a product of one MHV amplitude and one MHV form factor, and a single diagram containing the product $F^{(0)}_{n-1,1}\times A^{(0)}_{3,-1}$. For the former case one can use the BCFW bridge to rewrite the MHV$\times$MHV diagrams in terms of $R$-invariants. For the latter, instead, one has to use a recursive procedure. This results in the following combination of $(n-2)^2$ $R$-invariants:
\begin{equation}\label{genNHMV}
F^{(0)}_{n,1} \ = \ F^{(0)}_{n,0} \left(\sum_{j=3}^{n} \sum_{i=3}^{j} R'_{1;i,j}\, +\, \sum_{j=5}^{n+1}\sum_{i=3}^{j-2}  R''_{1;i,j} \right) \;,
\end{equation}
where the sum is meant to be periodic, {\it i.e.}~with the identification $n+1 \sim 1$. Notice that this representation has been obtained by using a particular BCFW shift, in this case $[1 \, 2\rangle$.

\section{Recursion relations for form factor integrands}
\label{recrelloop}
Given our prescription for the assignment of region variables in one-loop diagrams, we now proceed to consider the complete one-loop integrand $\mathcal{F}_{n,k}^{(1)}(x_0)$, defined by
\begin{equation}
 F_{n,k}^{(1)}=\int \!\mathrm{d}^d x_0 \; \mathcal{F}_{n,k}^{(1)}(\{x_i\};x_0) \ .
\end{equation}
In order to obtain recursion relations we perform particular
shifts of the external legs $\mathcal{F}_{n,k}^{(1)}(\{\hat{x}_i\};x_0)\equiv \widehat{\mathcal{F}}_{n,k}^{(1)}(z)$ such that 
\begin{equation}\label{restheor}
0 = \frac{1}{2 \pi \mathrm{i}} \oint \frac{\mathrm{d}z}{z} \; \widehat{\mathcal{F}}_{n,k}^{(1)}(z) = \mathcal{F}_{n,k}^{(1)}(\{x_i\};x_0) + \sum_{z_i \neq 0} \Res_{z=z_i} \frac{\widehat{\mathcal{F}}_{n,k}^{(1)}(z)}{z}
\ , 
\end{equation}
where the sum is taken over the residues of the integrand occurring at $z_i \neq 0$,  and we used $\widehat{\mathcal{F}}_{n,k}^{(1)}(0)=\mathcal{F}_{n,k}^{(1)}(\{x_i\};x_0)$.
Unitarity and locality guarantee that there are only first-order poles,  and we assume that the chosen deformation preserves the overall momentum conservation and leaves all particle momenta on shell. An important requirement is also that $ \widehat{\mathcal{F}}_{n,k}^{(1)}(z) \sim 1/z$ for large $z$. In this paper we only make use of deformations for which this is the case. We start with the case of a two-line shift, 
{\em i.e.}~the  loop-level generalisation of the familiar tree-level BCFW recursion relation.

\subsection{BCFW loop recursion relation}\label{SEC:BCFW}
In this section we consider a shift of the one-loop integrand that involves the shift of a single region momentum, together with all its periodic images. To be concrete, we focus on the shift
\begin{align}
\hat{x}^\bullet_1 \equiv x^\bullet_1-z \lambda_n \tilde \lambda_1 \;.
\end{align}
In terms of spinor variables, the above corresponds to a shift of the form
\begin{align}
\hat{\lambda}_1 &\equiv \lambda_1 - z \lambda_n \;, & \hat{\tilde{\lambda}}_n &\equiv \tilde{\lambda}_n + z \tilde{\lambda}_1 \;, & \hat{\eta}_n &\equiv \eta_n + z \eta_1 \;.
\end{align}

Similarly to the case of amplitudes, the residues of the integrand have simple physical origins: they are associated either to factorisation channels or to forward limits of tree-level form factors. 

As for form factors, with $\mathcal{A}^{(l)}_{n,k}$ we will denote the $l$-loop $n$-points $\mathrm{N}^k\mathrm{MHV}$ amplitude integrand. In our notation, with $l=0$ we simply denote the corresponding tree-level quantities. It is also useful to introduce the ratios defined by dividing form factors by the corresponding tree-level MHV quantities,
\begin{align}\label{ratios}
\tilde{\mathcal{F}}^{(l)}_{n,k}&\equiv \frac{\mathcal{F}^{(l)}_{n,k}}{\mathcal{F}^{(0)}_{n,0}} \;.
\end{align}

We can then propose the following formula for the one-loop integrand:
\begin{align}\label{recursion}
\mathcal{F}^{(1)}_{n,k} = \;
&F^{(0)}_{n,0} \; \tilde{\mathcal{F}}^{(1)}_{n-1,k}(\hat{x}_1,x_3,\dots, x_{n}, x_0) \cr
&+ \frac{1}{x_{01}^2} \, \int \mathrm{d}^4\eta_{\ell} \; \mathcal{F}_{n+2,k+1}^{(0)}(\hat{x}_1,\ldots,x_n,\hat{x}^-_1,x_0^-) \cr
&+\sum_{l,i,k_{\mathrm{L}}} \int \mathrm{d}^4\eta_{\ell} \; \bigg[ \mathcal{F}^{(l)}_{i,k_{\mathrm{L}}}(\hat{x}_1,\ldots,x_i) \; \frac{1}{(x_{i1}^+)^2} \; \mathcal{A}^{(1-l)}_{n-i+2,k_{\mathrm{R}}}(\hat{x}_1,x_i,\dots,x_n) \cr
&\kern5.6em + \mathcal{A}^{(l)}_{i,k_{\mathrm{L}}}(\hat{x}_1,\ldots,x_i) \; \frac{1}{(x_{i1})^2} \; \mathcal{F}^{(1-l)}_{n-i+2,k_{\mathrm{R}}}(\hat{x}_1,x_i,\dots,x_n) \bigg] \;,
\end{align}
where $l=0,1$, $i=2,\ldots,n-1$ and $k_{\mathrm{L}}+k_{\mathrm{R}}=k-1$ with $k_{\mathrm{L}}, k_{\mathrm{R}} \ge 0$, and $\eta_\ell$ is the Grassmann variable associated to the internal lines. We  will now systematically describe  the various terms in  this formula. Note  that for ease of notation we have dropped the dependence on $x_0$ in the last two lines.

The first line of \eqref{recursion} originates from the particular factorisation channel
\begin{align}\label{EQ:MHVb-NMHV_factorisation}
\raisebox{-.43\height}{
\begin{tikzpicture}[thick]
  \node[circle, fill=white, draw, minimum size=0.3cm] (L) at (-1.0,  0.0) {};
  \node[circle, fill=black!20, draw, minimum size=1.1cm] (R) at ( 1.0,  0.0) {};
  \node (Rlabel) at (R) {$\scriptstyle{\mathcal{F}^{(1)}_{n-1,k}}$};
  \draw (L) --node[circle, fill=white, draw=white]{} (R);
  \draw[red, dashed] (-0.17,0.26) node[black, anchor=south]{$x_3$} -- (-0.17,-0.26) node[black, anchor=north]{$\hat{x}_1$};
  \draw (L) -- ++(-120:0.8) node[anchor=east] {$\hat{1}$};
  \draw (L) -- ++( 120:0.8) node[anchor=east] {$2$};
  \draw (R) -- ++(  60:1.1) node[anchor=west] {$3$};
  \draw[dotted] (R)+(  50:0.9) to [bend left=30] ++(  10:0.9);
  \draw (R) -- ++(   0:1.1) node[anchor=west] {$\hat{n}$};
  \draw[double] (R) -- ++( -60:1.1) node[anchor=west] {};
\end{tikzpicture}
}
\qquad \longleftrightarrow \qquad
\raisebox{-.43\height}{
\begin{tikzpicture}[thick]
\coordinate (x1) at (0., 0.);
\coordinate (x1hat) at (-0.2, 0.4);
\coordinate (x2) at (0.5, 1.4);
\coordinate (x3) at (2.1, 1.);
\coordinate (xn) at (-1.5, -0.2);
\draw (xn) node [above left] {$x_n$};
\draw (x1) node [below] {$x_1$};
\draw (x1hat) node [above left] {$\hat{x}_1$};
\draw (x2) node [above] {$x_{2}$};
\draw (x3) node [above] {$x_{3}$};
\draw[ultra thick, blue!20] (xn) -- (x1) -- (x2);
\draw[ultra thick, blue] (x3) -- (x2) -- (x1hat) -- (xn);
\draw[wavy] (x1hat) -- (x3);
\end{tikzpicture}
}
\end{align}
which is the only one associated with the Parke--Taylor prefactor.
This diagram is evaluated in the particular kinematics for which $(\hat{x}_1-x_3)^2=0$. According to \eqref{ratios}, we can write the one-loop integrand in the above diagram as
\begin{align}
\mathcal{F}^{(1)}_{n-1,k} = F^{(0)}_{n-1,0} \; \tilde{\mathcal{F}}^{(1)}_{n-1,k}(\hat{x}_1,x_3,\dots, x_n;x_0) \;.
\end{align}
The tree-level prefactor recombines with the the $\overline{\mathrm{MHV}}$ amplitude, as in the BCFW recursion at tree level, to give the first line of \eqref{recursion}. Specifically,
\begin{align}
A_{3,-1}^{(0)} \, \frac{1}{x_{13}^2} \, \mathcal{F}^{(1)}_{n-1,k} &= A_{3,-1}^{(0)} \; \frac{1}{x_{13}^2} \; F^{(0)}_{n-1,0} \; \tilde{\mathcal{F}}^{(1)}_{n-1,k}(\hat{x}_1,x_3,\dots, x_n;x_0) \cr
&= F^{(0)}_{n,0} \; \tilde{\mathcal{F}}^{(1)}_{n-1,k}(\hat{x}_1,x_3,\dots, x_n;x_0) \;.
\end{align}

\begin{figure}[htb]
\[
\begin{tikzpicture}[thick,scale=0.9]
\path[use as bounding box] (-4.1, -4.9) rectangle (9.8, 5.7);
\begin{scope}[yshift=4.5cm, xshift=4cm, scale=0.7]
\drawULempty
\drawURempty
\drawLRempty
\drawLLempty
\draw (UL) -- (UR) -- (LR) --node[circle, fill=white, draw=white]{} (LL) --node[circle, fill=white, draw=white]{} (UL);
\draw (UL) -- ++( 150:1.0);
\draw[dotted] (UL)+( 145:0.8) to [bend left=30] ++( 125:0.8);
\draw (UL) -- ++( 120:1.0);
\draw (UR) -- ++(  30:1.0);
\draw[dotted] (UR)+(  55:0.8) to [bend left=30] ++(  35:0.8);
\draw (UR) -- ++(  60:1.0);
\draw (LR) -- ++( -30:1.0);
\draw[dotted] (LR)+( -35:0.8) to [bend left=30] ++( -55:0.8);
\draw (LR) -- ++( -60:1.0);
\draw[double] (LL) -- ++(-135:1.0);
\draw[red, dashed] (0,0)+(180:0.6) -- ++(180:1.1);
\draw[red, dashed] (0,0)+(-90:0.6) -- ++(-90:1.1);
\drawregionvariables{\raisebox{-0.6cm}{$\hat{x}_1^-$}}{$\hat{x}_1\quad$}{}{}{$x_0$}
\draw[green!50!black!50,->] (-2.2,0) to[out=180, in=90] (-8.5,-3);
\draw[green!50!black!50,->] (0,-2.2) to[out=-90, in=160] (2,-4);
\end{scope}
\begin{scope}[xshift=-3.5cm]
\coordinate (x1pre) at (0., 0.);
\coordinate (x1) at (0.5, 1.4);
\coordinate (x1hat) at (0.3, 1.9);
\coordinate (x1nxt) at (2.,1.);
\coordinate (x1preM) at ($(x1pre)+(3.0,0.)$);
\coordinate (x1M) at ($(x1)+(3.0,0.)$);
\coordinate (x1hatM) at ($(x1hat)+(3.0,0.)$);
\coordinate (x1nxtM) at ($(x1nxt)+(3.0,0.)$);
\coordinate (x0) at (2.2,2.);
\draw[ultra thick, blue!30] (x1pre) -- (x1) -- (x1nxt);
\draw[ultra thick, blue] (x1pre) -- (x1hat) -- (x1nxt);
\draw[ultra thick, blue, dotted] (x1nxt) -- (x1preM);
\draw[ultra thick, blue!30] (x1preM) -- (x1M) -- (x1nxtM);
\draw[ultra thick, blue] (x1preM) -- (x1hatM) -- (x1nxtM);
\draw (x1pre) node [below] {$x_{n}^{+}$};
\draw (x1) node [below right] {$\!\!x_{1}^{\vphantom{-}}$};
\draw (x1hat) node [above] {$\hat{x}_{1}$};
\draw (x1nxt) node [above right] {$x_{2}^{\vphantom{-}}$};
\draw (x1preM) node [below] {$x_{n}^{\vphantom{-}}$};
\draw (x1M) node [below right] {$\!\!x_{1}^-$};
\draw (x1hatM) node [above] {$\hat{x}_{1}^-$};
\draw (x1nxtM) node [above right] {$x_{2}^-$};
\draw (x0) node [above] {$x_{0}$};
\draw[wavy] (x0) -- (x1hat);
\end{scope}
\begin{scope}[xshift=3.5cm]
\coordinate (x1pre) at (0., 0.);
\coordinate (x1) at (0.5, 1.4);
\coordinate (x1hat) at (0.7, 1.);
\coordinate (x1nxt) at (2.,1.);
\coordinate (x1preM) at ($(x1pre)+(3.0,0.)$);
\coordinate (x1M) at ($(x1)+(3.0,0.)$);
\coordinate (x1hatM) at ($(x1hat)+(3.0,0.)$);
\coordinate (x1nxtM) at ($(x1nxt)+(3.0,0.)$);
\coordinate (x0) at (2.2,2.);
\draw[ultra thick, blue!30] (x1pre) -- (x1) -- (x1nxt);
\draw[ultra thick, blue] (x1pre) -- (x1hat) -- (x1nxt);
\draw[ultra thick, blue, dotted] (x1nxt) -- (x1preM);
\draw[ultra thick, blue!30] (x1preM) -- (x1M) -- (x1nxtM);
\draw[ultra thick, blue] (x1preM) -- (x1hatM) -- (x1nxtM);
\draw (x1pre) node [below] {$x_{n}^{+}$};
\draw (x1) node [above] {$x_{1}$};
\draw (x1hat) node [below right] {$\!\hat{x}_{1}^{\vphantom{-}}$};
\draw (x1nxt) node [right] {$\,\,x_{2}^{\vphantom{-}}$};
\draw (x1preM) node [below] {$x_{n}^{\vphantom{-}}$};
\draw (x1M) node [above] {$x_{1}^-$};
\draw (x1hatM) node [below right] {$\!\hat{x}_{1}^-$};
\draw (x1nxtM) node [right] {$\,\,x_{2}^-$};
\draw (x0) node [above] {$x_{0}$};
\draw[wavy] (x0) -- (x1hatM);
\end{scope}
\begin{scope}[yshift=-2.5cm, xshift=4cm, scale=0.7]
\drawULempty
\drawURempty
\drawLRempty
\drawLLempty
\draw (UL) -- (UR) -- (LR) --node[circle, fill=white, draw=white]{} (LL) --node[circle, fill=white, draw=white]{} (UL);
\draw (UL) -- ++( 150:1.0);
\draw[dotted] (UL)+( 145:0.8) to [bend left=30] ++( 125:0.8);
\draw (UL) -- ++( 120:1.0);
\draw (UR) -- ++(  30:1.0);
\draw[dotted] (UR)+(  55:0.8) to [bend left=30] ++(  35:0.8);
\draw (UR) -- ++(  60:1.0);
\draw (LR) -- ++( -30:1.0);
\draw[dotted] (LR)+( -35:0.8) to [bend left=30] ++( -55:0.8);
\draw (LR) -- ++( -60:1.0);
\draw[double] (LL) -- ++(-135:1.0);
\draw[red, dashed] (0,0)+(180:0.6) -- ++(180:1.1);
\draw[red, dashed] (0,0)+(-90:0.6) -- ++(-90:1.1);
\drawregionvariables{\raisebox{-0.6cm}{$\hat{x}_1^-$}}{$\hat{x}_1\quad$}{}{}{$x_0^-$}
\draw[green!50!black!50,->] (0,-2.2) to[out=-120, in=-90] (-8.5,3.5);
\end{scope}
\end{tikzpicture}
\]
\caption{The special kinematic configuration where $q$ is located between the shifted region momenta $\hat{x}_1$ and $\hat{x}_1^-$. The two corresponding residues reside on different periods of the periodic kinematic configuration. They can be mapped into each other by a shift of $x_0$. Note that for a generic configuration there is only a single residue as in the case of amplitudes.
}
\label{newdiags}
\end{figure}
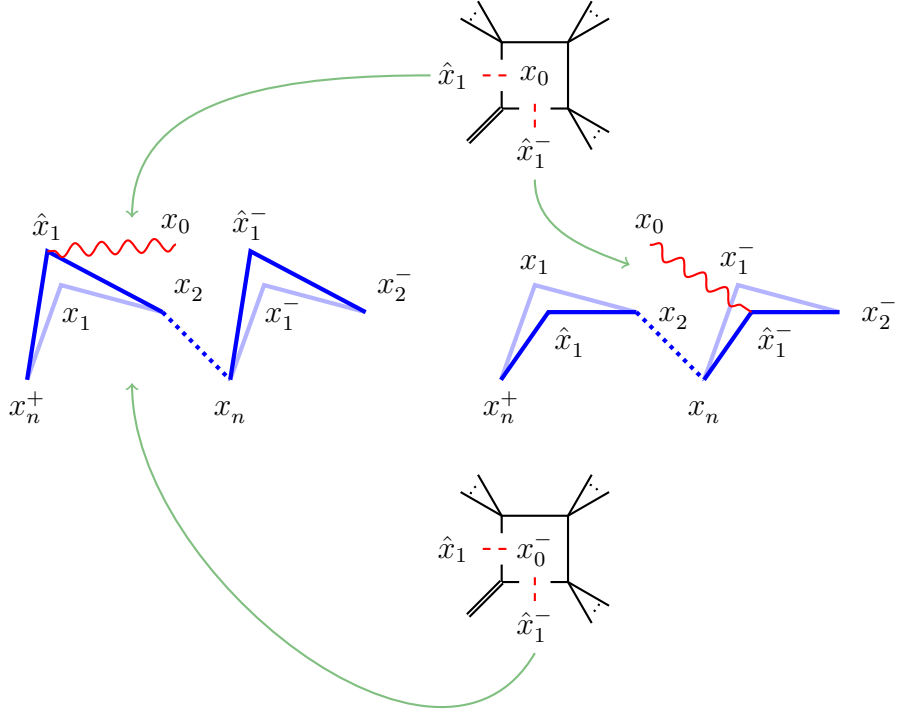

The second line of \eqref{recursion} contains the contributions from the forward limits. They are evaluated at the value of $z$ for which $(x_0-\hat{x}_{1})^2=0$. The geometric interpretation of the forward limit is shown in Figures~\ref{FIG:forward_limit} and \ref{FIG:single_cut}.
\begin{figure}[htb]
\begin{minipage}[b]{.5\linewidth}
\centering
\begin{tikzpicture}[thick]
\coordinate (x1) at (0.0, 0.);
\coordinate (x2) at (0.5, 1.);
\coordinate (x23) at (1.5, 0.8);
\coordinate (x3) at (2.3, 0.);
\coordinate (x4) at (2.4, 2.);
\coordinate (x1m) at ($(x1)+(2.5,0.)$);
\coordinate (x2m) at ($(x2)+(2.5,0.)$);
\coordinate (x23m) at ($(x23)+(2.5,0.)$);
\coordinate (x3m) at ($(x3)+(2.5,0.)$);
\coordinate (x4m) at ($(x4)+(2.5,0.)$);
\coordinate (x1mm) at ($(x1)+(5.0,0.)$);
\draw[ultra thick, blue] (x1) -- (x2) (x23) -- (x3) -- (x4) -- (x1m) -- (x2m) (x23m) -- (x3m) -- (x4m) -- (x1mm) ;
\draw[ultra thick, blue, dotted] (x1) -- ++(92:0.8) (x1mm) -- ++(65:0.8) (x2) -- (x23) (x2m) -- (x23m);
\draw (x1) node[below] {$x_{n+1}^+ \to x_1$};
\draw (x2) node[above] {$x_2$};
\draw (x23) node[above] {$x_n$};
\draw (x3) node[below] {$x_{n+1} \to x_1^-$};
\draw (x4) node[above] {$x_{n+2}$};
\draw (x2m) node[above] {$x_2^-$};
\draw (x23m) node[above] {$x_n$};
\draw (x3m) node[below] {$x_{n+1}^- \to x_1^{--}$};
\draw (x4m) node[above] {$x_{n+2}^-$};
\end{tikzpicture}
\subcaption{Forward limit}\label{FIG:forward_limit}
\end{minipage}
\begin{minipage}[b]{.5\linewidth}
\centering
\begin{tikzpicture}[thick]
\coordinate (x1) at (0.0, 0.);
\coordinate (x2) at (0.5, 1.);
\coordinate (x23) at (1.5, 0.8);
\coordinate (x0) at (2.4, 2.);
\coordinate (x1m) at ($(x1)+(2.5,0.)$);
\coordinate (x2m) at ($(x2)+(2.5,0.)$);
\coordinate (x23m) at ($(x23)+(2.5,0.)$);
\coordinate (x0m) at ($(x0)+(2.5,0.)$);
\coordinate (x1mm) at ($(x1)+(5.0,0.)$);
\draw[ultra thick, blue] (x1) -- (x2) (x23) -- (x1m) -- (x2m) (x23m) -- (x1mm);
\draw[ultra thick, blue, dotted] (x1) -- ++(150:0.8) (x1mm) -- ++(65:0.8) (x2) -- (x23) (x2m) -- (x23m);
\draw[wavy] (x0) -- (x1m);
\draw[wavy] (x0m) -- (x1mm);
\draw (x1) node[below] {$\hat{x}^{\phantom{-}}_1$};
\draw (x2) node[above] {$x_2$};
\draw (x23) node[above] {$x_n$};
\draw (x1m) node[below] {$\hat{x}_1^-$};
\draw (x0) node[above] {$x_0^-$};
\draw (x2m) node[above] {$x_2^-$};
\draw (x23m) node[above] {$x_n$};
\draw (x1mm) node[below] {$\hat{x}_1^{--}$};
\draw (x0m) node[above] {$x_0^{--}$};
\end{tikzpicture}
\subcaption{Single loop-leg cut}\label{FIG:single_cut}
\end{minipage}
\caption{Illustration of the forward limits and single cuts on the periodic kinematic configuration. The red wiggly represents
the distance $\hat{x}_{01}$ that becomes null at the location of the residue. This also explains the arguments of the $(n+2)$-point form factor
appearing in the second line of \eqref{recursion}.}
\end{figure}
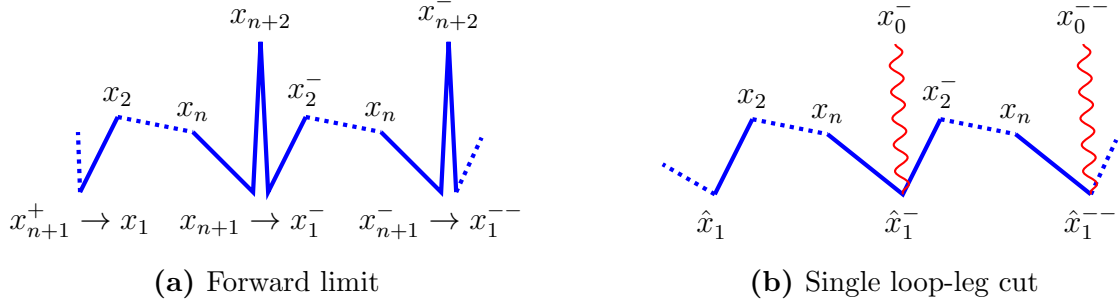\\
Compared to the recursion relation of amplitude integrands, there is an important difference arising from diagrams where the shifted region variable $x_i$ appears twice in the expression of the integrand (see Figure~\ref{newdiags}). This occurs when the operator carrying momentum $q$ is located between the shifted region momenta $\hat{x}_1$ and $\hat{x}_1^-$. 
When taking the sum over the residues, these diagrams will give two contributions: one arising from a pole when $\hat{x}_{01}^2=0$, and another one from a pole at  $(\hat{x}_{01}^+)^2=0$. These two poles are associated with two different values of~$z$. However, as discussed in Section~\ref{regionvar}, we can use the freedom of shifting $x_0$ by a period to find a representation of the integrand with an overall factor $1/x_{01}^2$. Notice that, since $z$ itself depends on $x_0$, this gets shifted as well and the two residues are then associated with the same value of $z$. One may still wonder whether both these contributions are produced in the forward limit of some higher point amplitudes; this is indeed the case and we will demonstrate this in specific examples later on. 

\newpage

Finally, in the last two lines of \eqref{recursion} every pole is associated with a standard factorisation channel, thus $z$ is evaluated respectively at $(x_i-\hat{x}_1^-)^2=0$ and $(x_i-\hat{x}_1)^2=0$ as illustrated below:
\begin{align}
\raisebox{-.45\height}{
\begin{tikzpicture}[thick, scale=1.3]
  \node[circle, fill=black!20, draw, minimum size=0.9cm] (L) at (-1.0,  0.0) {};
  \node[circle, fill=black!20, draw, minimum size=0.9cm] (R) at ( 1.0,  0.0) {};
  \node (Rlabel) at (L) {$\scriptstyle{\mathcal{F}_{\mathrm{L}}}$};
  \node (Rlabel) at (R) {$\scriptstyle{\mathcal{A}_{\mathrm{R}}}$};
  \draw (L) --node[circle, fill=white, draw=white]{} (R);
  \draw[red, dashed] (0.0,0.21) node[black, anchor=south]{$x_i$} -- (0.0,-0.21) node[black, anchor=north]{$\hat{x}_1^-$};
  \draw[double] (L) -- ++(-120:0.8);
  \draw (L) -- ++(-180:0.8) node[left] {$\hat{1}$};
  \draw[dotted] (L)+( 130:0.6) to [bend right=30] ++( 170:0.6);
  \draw (L) -- ++( 120:0.8) node[left] {$i-1$};
  \draw (R) -- ++(  60:0.8) node[right] {$i$};
  \draw[dotted] (R)+(  50:0.6) to [bend left=60] ++( -50:0.6);
  \draw (R) -- ++( -60:0.8) node[right] {$\hat{n}$};
\end{tikzpicture}
}
\kern 1.25em \quad&\longleftrightarrow\quad
\raisebox{-.43\height}{
\begin{tikzpicture}[ultra thick, scale=1.6]
  \coordinate (xnP) at (-0.4, 0.7);
  \coordinate (x1o) at ( 0.0, 0.45);
  \coordinate (x1)  at (-0.2, 0.0);
  \coordinate (x2)  at ( 0.4, 0.4);
  \coordinate (xim) at ( 1.0, 0.6);
  \coordinate (xi)  at ( 1.4, 0.3);
  \coordinate (xip) at ( 1.7, 0.8);
  \coordinate (xn) at  ( 2.4, 0.7);
  \coordinate (x1M) at ( 2.6, 0.0);
  \coordinate (x1oM) at( 2.8, 0.45);
  \coordinate (x2M) at ( 3.2, 0.4);
  \draw[blue!20] (xnP) -- (x1o) -- (x2) (xn) -- (x1oM) -- (x2M);
  \draw[blue] (xnP) -- (x1) -- (x2) (xim) -- (xi) -- (xip) (xn) -- (x1M) -- (x2M);
  \draw[blue, dotted] (x2) -- (xim) (xip) -- (xn);
  \draw[wavy, thick] (x1M) -- (xi);
  \draw (xnP) node[above] {$x_n^+$};
  \draw (x1) node[below] {$\hat{x}_1$};
  \draw (x1o) node[above] {$x_1$};
  \draw (x2) node[above] {$x_2$};
  \draw (xim) node[above] {$x_{i-1}$};
  \draw (xi) node[below] {$x_i$};
  \draw (xip) node[above] {$x_{i+1}$};
  \draw (xn) node[above] {$x_n$};
  \draw (x1M) node[below] {$\hat{x}_1^-$};
  \draw (x1oM) node[above] {$x^-_1$};
  \draw (x2M) node[above] {$x_2^-$};
\end{tikzpicture}
}
\cr
\raisebox{-.45\height}{
\begin{tikzpicture}[thick, scale=1.3]
  \node[circle, fill=black!20, draw, minimum size=0.9cm] (L) at (-1.0,  0.0) {};
  \node[circle, fill=black!20, draw, minimum size=0.9cm] (R) at ( 1.0,  0.0) {};
  \node (Rlabel) at (L) {$\scriptstyle{\mathcal{A}_{\mathrm{L}}}$};
  \node (Rlabel) at (R) {$\scriptstyle{\mathcal{F}_{\mathrm{R}}}$};
  \draw (L) --node[circle, fill=white, draw=white]{} (R);
  \draw[red, dashed] (0.0,0.21) node[black, anchor=south]{$x_i$} -- (0.0,-0.21) node[black, anchor=north]{$\hat{x}_1$};
  \draw (L) -- ++(-120:0.8) node[left] {$\hat{1}$};
  \draw[dotted] (L)+( 130:0.6) to [bend right=60] ++( 230:0.6);
  \draw (L) -- ++( 120:0.8) node[left] {$i-1$};
  \draw (R) -- ++(  60:0.8) node[right] {$i$};
  \draw[dotted] (R)+(  50:0.6) to [bend left=30] ++(  10:0.6);
  \draw (R) -- ++(   0:0.8) node[right] {$\hat{n}$};
  \draw[double] (R) -- ++( -60:0.8) node[right] {};
\end{tikzpicture}
}
\quad&\longleftrightarrow\quad
\raisebox{-.43\height}{
\begin{tikzpicture}[ultra thick, scale=1.6]
  \coordinate (xnP) at (-0.4, 0.7);
  \coordinate (x1o) at ( 0.0, 0.45);
  \coordinate (x1)  at ( 0.2, 0.0);
  \coordinate (x2)  at ( 0.4, 0.4);
  \coordinate (xim) at ( 1.0, 0.6);
  \coordinate (xi)  at ( 1.4, 0.3);
  \coordinate (xip) at ( 1.7, 0.8);
  \coordinate (xn) at  ( 2.4, 0.7);
  \coordinate (x1M) at ( 3.0, 0.0);
  \coordinate (x1oM) at( 2.8, 0.45);
  \coordinate (x2M) at ( 3.2, 0.4);
  \draw[blue!20] (xnP) -- (x1o) -- (x2) (xn) -- (x1oM) -- (x2M);
  \draw[blue] (xnP) -- (x1) -- (x2) (xim) -- (xi) -- (xip) (xn) -- (x1M) -- (x2M);
  \draw[blue, dotted] (x2) -- (xim) (xip) -- (xn);
  \draw[wavy, thick] (x1) -- (xi);
  \draw (xnP) node[above] {$x_n^+$};
  \draw (x1) node[below] {$\hat{x}_1$};
  \draw (x1o) node[above] {$x_1$};
  \draw (x2) node[above] {$x_2$};
  \draw (xim) node[above] {$x_{i-1}$};
  \draw (xi) node[below] {$x_i$};
  \draw (xip) node[above] {$x_{i+1}$};
  \draw (xn) node[above] {$x_n$};
  \draw (x1M) node[below] {$\hat{x}_1^-$};
  \draw (x1oM) node[above] {$x^-_1$};
  \draw (x2M) node[above] {$x_2^-$};
\end{tikzpicture}
}
\end{align}

Given the one-loop recursion relation \eqref{recursion}, it is tempting to propose at this point a straightforward all-loop generalisation:
\begin{align}\label{alllooprecursion}
\mathcal{F}^{(l)}_{n,k} = \;
&F^{(0)}_{n,k} \; \tilde{\mathcal{F}}^{(l)}_{n-1,k}(\hat{x}_1,x_3,\dots, x_n, x_0) \cr
&+ \frac{1}{x_{01}^2} \, \int \mathrm{d}^4\eta_{\ell} \; \mathcal{F}_{n+2,k+1}^{(l-1)}(\hat{x}_1,\ldots,x_n,\hat{x}^-_1,x_0^-) \cr
&+\sum_{l_{\mathrm{L}},i,k_{\mathrm{L}}}  \int \mathrm{d}^4\eta_{\ell} \; \bigg[\mathcal{F}^{(l_{\mathrm{L}})}_{i,k_{\mathrm{L}}}(\hat{x}_1,\ldots,x_i) \; \frac{1}{(x_{i1}^+)^2} \; \mathcal{A}^{(l_{\mathrm{R}})}_{n-i+2,k_{\mathrm{R}}}(\hat{x}_1,x_i,\dots,x_n) \cr
&\kern6.1em + \mathcal{A}^{(l_{\mathrm{L}})}_{i,k_{\mathrm{L}}}(\hat{x}_1,\ldots,x_i) \; \frac{1}{(x_{i1})^2} \; \mathcal{F}^{(l_{\mathrm{R}})}_{n-i+2,k_{\mathrm{R}}}(\hat{x}_1,x_i,\dots,x_n) \Big] \;,
\end{align}
with $l_{\mathrm{L}}+l_{\mathrm{R}}=l$, $i=2,\dots n-1$ and $k_{\mathrm{L}}+k_{\mathrm{R}}=k-1$ with $k_{\mathrm{L}},k_{\mathrm{R}} \ge 0$. In this expression we
have suppressed lower loop variables for easy of notation and we only quote $x_0$ corresponding to the new variable.
One of the issues that needs to be clarified at higher loops is the assignment of region variables and the associated ambiguity we discussed in Section \ref{regionvar}. We leave this analysis for the future, and in this work we focus on  explicit checks of the one-loop recursion presented in \eqref{recursion}.

The examples we discuss in the following are one-loop MHV form factors. In this case, the recursion has only two contributions:
\begin{align}
\mathcal{F}^{(1)}_{n,k} = F^{(0)}_{n,0} \; \tilde{\mathcal{F}}^{(1)}_{n-1,k}(\hat{x}_1,x_3,\dots, x_n, x_0) + \frac{1}{x_{01}^2} \, \int \mathrm{d}^4\eta_{\ell} \; \mathcal{F}_{n+2,k+1}^{(0)}(\hat{x}_1,\ldots,x_n,\hat{x}^-_1,x_0^-) \;.
\end{align}

In the next two subsections we provide examples of the BCFW recursion at one loop. We show the validity of our approach by comparing  results obtained  by using our prescription in \eqref{recursion} with integrands obtained from generalised unitarity. To show agreement between the two, we will explicitly check that the result obtained with unitarity methods has residues coming from single loop-leg cuts which are precisely captured by forward limits of tree-level form factors, up to shifts in $x_0$.

\subsubsection{The one-loop two-point form factor}
The simplest example is given by the minimal form factor at one loop. As anticipated, we start by considering the one-loop integrand coming from generalised unitarity. In this case only triangles can appear. When summing over cyclic permutations of the external on-shell legs, one obtains
\begin{align}\label{EQ:integrand_F2,0}
\mathcal{F}^{(1)}_{2,0}(x_1,x_2;x_0) = F^{(0)}_{2,0} \left(s_{12}
\raisebox{-.47\height}{
\begin{tikzpicture}[thick, scale=0.6]
\coordinate (T1) at ( 0.5 ,  0.87);
\coordinate (T2) at ( 0.5 , -0.87);
\coordinate (T3) at (-1.0 ,  0.0 );
\draw (T1) -- (T2) -- (T3) -- (T1);
\node[] (Tlbl3) at (-0.6 ,  0.8) {$x_1$};
\node[] (Tlbl1) at ( 1.2 ,  0.0) {$x_2$};
\node[] (Tlbl2) at (-0.6 , -0.8) {$x_1^-$};
\node[] (Tlbl0) at ( 0.0 ,  0.0) {$x_0$};
\draw (T1) -- ++(0:0.6) node[right] {};
\draw (T2) -- ++(0:0.6) node[right] {};
\draw[double] (T3) -- ++(180:0.6) node[left] {};
\end{tikzpicture}
} + \; s_{12}
\raisebox{-.47\height}{
\begin{tikzpicture}[thick, scale=0.6]
\coordinate (T1) at ( 0.5,  0.87);
\coordinate (T2) at ( 0.5, -0.87);
\coordinate (T3) at (-1.0,  0.0 );
\draw (T1) -- (T2) -- (T3) -- (T1);
\node[] (Tlbl3) at (-0.6,  0.8) {$x_2$};
\node[] (Tlbl1) at ( 1.2,  0.0) {$x_1^-$};
\node[] (Tlbl2) at (-0.6, -0.8) {$x_2^-$};
\node[] (Tlbl0) at ( 0.0,  0.0) {$x_0$};
\draw (T1) -- ++(0:0.6) node[right] {};
\draw (T2) -- ++(0:0.6) node[right] {};
\draw[double] (T3) -- ++(180:0.6) node[left] {};
\end{tikzpicture}
} \right) \;.
\end{align}
We now consider the BCFW shift
\begin{align}
\hat{x}_{1}^\bullet \equiv x_1^\bullet - z \lambda_{2}\widetilde{\lambda}_{1} \;, 
\end{align}
and collect all the residues associated with it. These come from the cuts
\begin{align}
\raisebox{-.47\height}{
\begin{tikzpicture}[thick, scale=0.75]
\coordinate (T1) at ( 0.5 ,  0.87);
\coordinate (T2) at ( 0.5 , -0.87);
\coordinate (T3) at (-1.0 ,  0.0 );
\draw (T1) -- (T2) -- (T3) --node[circle, fill=white, draw=white]{} (T1);
\draw[red, dashed] (0,0)+(120:0.3) -- ++(120:0.7);
\node[] (Tlbl3) at (-0.6 ,  0.8) {$\hat{x}_1$};
\node[] (Tlbl1) at ( 1.2 ,  0.0) {$x_2$};
\node[] (Tlbl2) at (-0.6 , -0.8) {$\hat{x}_1^-$};
\node[] (Tlbl0) at ( 0.0 ,  0.0) {$x_0$};
\draw (T1) -- ++(0:0.6); 
\draw (T2) -- ++(0:0.6); 
\draw[double] (T3) -- ++(180:0.6) node[left] {};
\end{tikzpicture}
} &= -\frac{1}{x_{01}^2 x_{02}^2 (\hat{x}_{01}^+)^2} \;, &
\raisebox{-.47\height}{
\begin{tikzpicture}[thick, scale=0.75]
\coordinate (T1) at ( 0.5 ,  0.87);
\coordinate (T2) at ( 0.5 , -0.87);
\coordinate (T3) at (-1.0 ,  0.0 );
\draw (T1) -- (T2) --node[circle, fill=white, draw=white]{} (T3) -- (T1);
\draw[red, dashed] (0,0)+(240:0.3) -- ++(240:0.7);
\node[] (Tlbl3) at (-0.6 ,  0.8) {$\hat{x}_1$};
\node[] (Tlbl1) at ( 1.2 ,  0.0) {$x_2$};
\node[] (Tlbl2) at (-0.6 , -0.8) {$\hat{x}_1^-$};
\node[] (Tlbl0) at ( 0.0 ,  0.0) {$x_0$};
\draw (T1) -- ++(0:0.6); 
\draw (T2) -- ++(0:0.6); 
\draw[double] (T3) -- ++(180:0.6) node[left] {};
\end{tikzpicture}
} &= -\frac{1}{(x_{01}^+)^2 \hat{x}_{01}^2 x_{02}^2} \;, \cr
\raisebox{-.47\height}{
\begin{tikzpicture}[thick, scale=0.75]
\coordinate (T1) at ( 0.5 ,  0.87);
\coordinate (T2) at ( 0.5 , -0.87);
\coordinate (T3) at (-1.0 ,  0.0 );
\draw (T1) --node[circle, fill=white, draw=white]{} (T2) -- (T3) -- (T1);
\draw[red, dashed] (0,0)+(  0:0.3) -- ++(  0:0.7);
\node[] (Tlbl3) at (-0.6,  0.8) {$x_2$};
\node[] (Tlbl1) at ( 1.2,  0.0) {$\hat{x}_1^-$};
\node[] (Tlbl2) at (-0.6, -0.8) {$x_2^-$};
\node[] (Tlbl0) at ( 0.0,  0.0) {$x_0$};
\draw (T1) -- ++(0:0.6); 
\draw (T2) -- ++(0:0.6); 
\draw[double] (T3) -- ++(180:0.6) node[left] {};
\end{tikzpicture}
} &= -\frac{1}{(x_{01}^+)^2 x_{02}^2(x_{02}^+)^2} \;. & &
\end{align}
Similarly to the situation depicted in Figure~\ref{newdiags}, the first triangle in \eqref{EQ:integrand_F2,0} gives rise to two different cut contributions.
Notice also how, in this particular case, the triangle coefficients and the MHV prefactor are insensitive to the deformation.
Let us collect the three terms in a single function $\mathcal{I}_{\mathrm{cut}}$. In doing so, we perform the shift $x_0\mapsto x_0^-$ on the last two terms to obtain a universal prefactor $1/(x_{01})^{2}$ associated with the cut leg. Hence, we obtain
\begin{align}
\mathcal{I}_{\mathrm{cut}} = -\frac{F^{(0)}_{2,0}}{x_{01}^2}\left(\frac{s_{12}}{x_{02}^2 (\hat{x}_{01}^+)^2} + \frac{s_{12}}{(\hat{x}_{01}^-)^2 (x_{02}^-)^2} + \frac{s_{12}}{(x_{02}^-)^2x_{02}^2}\right) \;.
\end{align}
According to \eqref{recursion}, we can reproduce the results above from the forward limit of $F^{(0)}_{4,1}$. The expression for this, given in \eqref{genNHMV}, reads
\begin{align}
\tilde{F}^{(0)}_{4,1} = R'_{133}+R'_{134}+R'_{144}+R''_{131} \;.
\end{align}
When taking the forward limit, we make the assignments
\begin{align}\label{forward4pt}
\lambda_4&\to-\lambda_3 \:, & \widetilde{\lambda}_4&\to\widetilde{\lambda}_3 \;, & \eta_4&=\eta_3 \; .
\end{align}
By looking at the expressions of the $R$-invariants it is easy to see that some of the denominators vanish under these identifications. In particular, this happens when legs $3$ and $4$ are attached to the same MHV blob as in the first two diagrams in the second line of \eqref{EQ:BCFW_NMHV} for $n=4$, \textit{i.e.}\ $R'_{133}$ and $R''_{131}$. Similar diagrams were already considered in the amplitude case \cite{Brandhuber:2005kd,CaronHuot:2010zt, ArkaniHamed:2010kv}. It turns out that for supersymmetric theories their contribution vanishes in the sum over all the possible external states appearing in the two legs with momenta $p_3$ and $-p_3$ in the forward limit. For $\mathcal{N}=4$ SYM the sum over the states can be implemented by integrating over the Grassmann variable $\eta_3$. Looking at the expressions for the $R$-invariants $R'_{133}$ and $R''_{131}$ one immediately notices that the dependence on $\eta_4$ disappears in the configuration \eqref{forward4pt}. This implies that the integration over $\eta_3$ will always vanish when legs $3$ and $4$ are attached to the same MHV blob. This provides a  systematic and graphical way to isolate the diagrams that contribute to the forward limit of the NMHV amplitude.
Therefore, after integrating over $\mathrm{d}^4\eta_3$ we are left with the following contributions:
\begin{align}
\begin{split}
\lim_{p_4 \rightarrow -p_3} \int \mathrm{d}^4\eta_3 \, R'_{134} &= \frac{\delta^{(8)}(\mathrm{q}) \, [12]^2}{(p_3-q)^2 \, (p_3-p_1)^2 \, 2p_3\cdot q} 
\\
&=
\frac{\delta^{(8)}(\mathrm{q}) \, [12]^2}{q^2 \,  (p_3-p_1)^2 \, 2 p_3 \cdot
q}
+ \frac{\delta^{(8)}(\mathrm{q}) \, [12]^2}{q^2 \, (p_3-q)^2 \, (p_3-p_1)^2} \;, 
\end{split}
\end{align}
\begin{align}
\begin{split}
\lim_{p_4 \rightarrow -p_3} \int \mathrm{d}^4\eta_3 \,  R'_{144} &= -\frac{\delta^{(8)}(\mathrm{q}) \, [12]^2}{(p_3+q)^2 \, (p_3+p_2)^2 \, 2p_3 \cdot q}
\\ &
= -\frac{\delta^{(8)}(\mathrm{q}) \, [12]^2}{q^2 \, (p_3+p_2)^2 \, 2p_3 \cdot q} + \frac{\delta^{(8)}(\mathrm{q}) \, [12]^2}{q^2 \, (p_3+q)^2 \, (p_3+p_2)^2} \;.
\end{split}
\end{align}
The sum of these expressions gives
\begin{align}\label{F4_forward}
\mathcal{I}_{\mathrm{forw}} &= - F^{(0)}_{2,0} \left(\frac{s_{12}}{s_{1,-3} s_{1,2,-3}} + \frac{s_{12}}{s_{1,2,3} \, s_{2,3}} + \frac{s_{12}}{s_{2,3} s_{1,-3}}\right) \;,
\end{align}
where we used the notation $s_{i,\dots, \pm j}=(p_i+\dots \pm p_j)^2$. This is the result for the forward limit. According to \eqref{recursion} we need to evaluate this expression on a shifted kinematics. First we express Mandelstam variables in terms of  region variables, using the forward kinematics $x_3 = x_1^-$, $x_4 = x_0^-$. Then we simply shift $x_1 \mapsto \hat{x}_1$ as shown in Figure~\ref{FIG:forward_limit}.

With this, we obtain full agreement between the two expressions, as stated in \eqref{recursion}, {\it i.e.}\ we have
\begin{align}
\mathcal{I}_{\mathrm{cut}} = \frac{1}{x_{01}^2} \, \hat{\mathcal{I}}_{\mathrm{forw}} \;,
\end{align}
where in $\hat{\mathcal{I}}_{\mathrm{forw}}$ we performed the identification described above.
In this particular case, $\mathcal{I}_{\mathrm{cut}}$ reconstructs the full integrand, since, as noted earlier, the poles we considered are all the poles of the integrand function.

\subsubsection{The one-loop three-point MHV form factor}
The example described in the previous section is very simple because of the small number of diagrams and the absence of boxes. The first case where box diagrams appear is the three-point case, whose one-loop integrand was derived in \cite{Brandhuber:2010ad}. This result, expressed using the region variable assignment described in Section~\ref{regionvar}, reads
\begin{align}
\frac{\mathcal{F}^{(1)}_{3,1}(x_0)}{F^{(0)}_{3,0}} = \;&\frac{x_{13}^2 (x_{21}^+)^2}{2}
\raisebox{-0.47\height}{
\begin{tikzpicture}[thick,scale=0.5]
\drawULempty
\drawURempty
\drawLRempty
\drawLLempty
\draw (UL) -- (UR) -- (LR) -- (LL) -- (UL);
\draw[double] (UL) -- ++( 135:1.0);
\draw (UR) -- ++(  45:1.0) node[anchor=south west] {};
\draw (LR) -- ++( -45:1.0) node[anchor=north west] {};
\draw (LL) -- ++(-135:1.0) node[anchor=north east] {};
\drawregionvariables{$x_3$}{$x_1^-\;$}{$x_1$}{$\; x_2$}{$x_0$}
\end{tikzpicture}
}
+ \frac{(x_{21}^+)^2 (x_{32}^+)^2}{2}
\raisebox{-0.48\height}{
\begin{tikzpicture}[thick,scale=0.5]
\drawULempty
\drawURempty
\drawLRempty
\drawLLempty
\draw (UL) -- (UR) -- (LR) -- (LL) -- (UL);
\draw[double] (UL) -- ++( 135:1.0);
\draw (UR) -- ++(  45:1.0) node[anchor=south west] {};
\draw (LR) -- ++( -45:1.0) node[anchor=north west] {};
\draw (LL) -- ++(-135:1.0) node[anchor=north east] {};
\drawregionvariables{\raisebox{-0.4cm}{$x_1^-$}}{$x_2^-\;$}{$x_2$}{$\; x_3$}{$x_0$}
\end{tikzpicture}
}
+ \frac{(x_{32}^+)^2 x_{13}^2}{2}
\raisebox{-0.48\height}{
\begin{tikzpicture}[thick,scale=0.5]
\drawULempty
\drawURempty
\drawLRempty
\drawLLempty
\draw (UL) -- (UR) -- (LR) -- (LL) -- (UL);
\draw[double] (UL) -- ++( 135:1.0);
\draw (UR) -- ++(  45:1.0) node[anchor=south west] {};
\draw (LR) -- ++( -45:1.0) node[anchor=north west] {};
\draw (LL) -- ++(-135:1.0) node[anchor=north east] {};
\drawregionvariables{\raisebox{-0.4cm}{$x_2^-$}}{$x_3^-\;$}{$x_3$}{$\; x_1^-$}{$x_0$}
\end{tikzpicture}
} \cr
&+ \frac{(x_{21}^+)^2+(x_{32}^+)^2}{2} \left(
\raisebox{-.45\height}{
\begin{tikzpicture}[thick, scale=0.6]
\coordinate (T1) at ( 0.5 ,  0.87);
\coordinate (T2) at ( 0.5 , -0.87);
\coordinate (T3) at (-1.0 ,  0.0 );
\draw (T1) -- (T2) -- (T3) -- (T1);
\node[] (Tlbl3) at (-0.6,  0.8) {$x_1$};
\node[] (Tlbl1) at ( 1.2,  0.0) {$x_3$};
\node[] (Tlbl2) at (-0.6, -0.8) {$x_1^-$};
\node[] (Tlbl0) at ( 0.0,  0.0) {$x_0$};
\draw (T1) -- ++( 20:0.6) node[anchor=west] {};
\draw (T1) -- ++(-20:0.6) node[anchor=west] {};
\draw (T2) -- ++(0:0.6) node[anchor=west] {};
\draw[double] (T3) -- ++(180:0.6) node[anchor=east] {};
\end{tikzpicture}
} + 
\raisebox{-.45\height}{
\begin{tikzpicture}[thick, scale=0.6]
\coordinate (T1) at ( 0.5 ,  0.87);
\coordinate (T2) at ( 0.5 , -0.87);
\coordinate (T3) at (-1.0 ,  0.0 );
\draw (T1) -- (T2) -- (T3) -- (T1);
\node[] (Tlbl3) at (-0.6,  0.8) {$x_3$};
\node[] (Tlbl1) at ( 1.2,  0.0) {$x_1^-$};
\node[] (Tlbl2) at (-0.6, -0.8) {$x_3^-$};
\node[] (Tlbl0) at ( 0.0,  0.0) {$x_0$};
\draw (T1) -- ++(0:0.6) node[anchor=west] {};
\draw (T2) -- ++( 20:0.6) node[anchor=west] {};
\draw (T2) -- ++(-20:0.6) node[anchor=west] {};
\draw[double] (T3) -- ++(180:0.6) node[anchor=east] {};
\end{tikzpicture}
} \right) \cr
&+ \kern 1.2em \frac{(x_{32}^+)^2+x_{13}^2}{2} \left(
\raisebox{-.45\height}{
\begin{tikzpicture}[thick, scale=0.6]
\coordinate (T1) at ( 0.5 ,  0.87);
\coordinate (T2) at ( 0.5 , -0.87);
\coordinate (T3) at (-1.0 ,  0.0 );
\draw (T1) -- (T2) -- (T3) -- (T1);
\node[] (Tlbl3) at (-0.6,  0.8) {$x_2$};
\node[] (Tlbl1) at ( 1.2,  0.0) {$x_1^-$};
\node[] (Tlbl2) at (-0.6, -0.8) {$x_2^-$};
\node[] (Tlbl0) at ( 0.0,  0.0) {$x_0$};
\draw (T1) -- ++( 20:0.6) node[anchor=west] {};
\draw (T1) -- ++(-20:0.6) node[anchor=west] {};
\draw (T2) -- ++(0:0.6) node[anchor=west] {};
\draw[double] (T3) -- ++(180:0.6) node[anchor=east] {};
\end{tikzpicture}
} + 
\raisebox{-.45\height}{
\begin{tikzpicture}[thick, scale=0.6]
\coordinate (T1) at ( 0.5 ,  0.87);
\coordinate (T2) at ( 0.5 , -0.87);
\coordinate (T3) at (-1.0 ,  0.0 );
\draw (T1) -- (T2) -- (T3) -- (T1);
\node[] (Tlbl3) at (-0.6,  0.8) {$x_1$};
\node[] (Tlbl1) at ( 1.2,  0.0) {$x_2$};
\node[] (Tlbl2) at (-0.6, -0.8) {$x_1^-$};
\node[] (Tlbl0) at ( 0.0,  0.0) {$x_0$};
\draw (T1) -- ++(0:0.6) node[anchor=west] {};
\draw (T2) -- ++( 20:0.6) node[anchor=west] {};
\draw (T2) -- ++(-20:0.6) node[anchor=west] {};
\draw[double] (T3) -- ++(180:0.6) node[anchor=east] {};
\end{tikzpicture}
} \right) \cr
&+ \kern 1.2em \frac{x_{13}^2 +(x_{21}^+)^2}{2} \left(
\raisebox{-.45\height}{
\begin{tikzpicture}[thick, scale=0.6]
\coordinate (T1) at ( 0.5 ,  0.87);
\coordinate (T2) at ( 0.5 , -0.87);
\coordinate (T3) at (-1.0 ,  0.0 );
\draw (T1) -- (T2) -- (T3) -- (T1);
\node[] (Tlbl3) at (-0.6,  0.8) {$x_3$};
\node[] (Tlbl1) at ( 1.2,  0.0) {$x_2^-$};
\node[] (Tlbl2) at (-0.6, -0.8) {$x_3^-$};
\node[] (Tlbl0) at ( 0.0,  0.0) {$x_0$};
\draw (T1) -- ++( 20:0.6) node[anchor=west] {};
\draw (T1) -- ++(-20:0.6) node[anchor=west] {};
\draw (T2) -- ++(0:0.6) node[anchor=west] {};
\draw[double] (T3) -- ++(180:0.6) node[anchor=east] {};
\end{tikzpicture}
} + 
\raisebox{-.45\height}{
\begin{tikzpicture}[thick, scale=0.6]
\coordinate (T1) at ( 0.5 ,  0.87);
\coordinate (T2) at ( 0.5 , -0.87);
\coordinate (T3) at (-1.0 ,  0.0 );
\draw (T1) -- (T2) -- (T3) -- (T1);
\node[] (Tlbl3) at (-0.6,  0.8) {$x_2$};
\node[] (Tlbl1) at ( 1.2,  0.0) {$x_3$};
\node[] (Tlbl2) at (-0.6, -0.8) {$x_2^-$};
\node[] (Tlbl0) at ( 0.0,  0.0) {$x_0$};
\draw (T1) -- ++(0:0.6) node[anchor=west] {};
\draw (T2) -- ++( 20:0.6) node[anchor=west] {};
\draw (T2) -- ++(-20:0.6) node[anchor=west] {};
\draw[double] (T3) -- ++(180:0.6) node[anchor=east] {};
\end{tikzpicture}
} \right) \;.
\end{align}
We then consider the BCFW shift
\begin{equation}
\hat{x}^\bullet_1 \equiv x^\bullet_1 - z \lambda_3 \tilde{\lambda}_1 \ , 
\end{equation}
and collect the residues coming from the the above expression. These are associated with the cuts
\begin{align}
\raisebox{-0.47\height}{
\begin{tikzpicture}[thick,scale=0.6]
\drawULempty
\drawURempty
\drawLRempty
\drawLLempty
\draw (UL) --node[circle, fill=white, draw=white]{} (UR) -- (LR) -- (LL) -- (UL);
\draw[double] (UL) -- ++( 135:1.0);
\draw (UR) -- ++(  45:1.0) node[anchor=south west] {};
\draw (LR) -- ++( -45:1.0) node[anchor=north west] {};
\draw (LL) -- ++(-135:1.0) node[anchor=north east] {};
\draw[red, dashed] (0,0)+( 90:0.5) -- ++( 90:1.0);
\drawregionvariables{$x_3$}{$\hat{x}_1^-\;$}{$\hat{x}_1$}{$\; x_2$}{$x_0$}
\end{tikzpicture}
} &= \frac{1}{x_{01}^2 x_{02}^2 x_{03}^2 (\hat{x}_{01}^+)^2} \;, &
\raisebox{-0.47\height}{
\begin{tikzpicture}[thick,scale=0.6]
\drawULempty
\drawURempty
\drawLRempty
\drawLLempty
\draw (UL) -- (UR) -- (LR) -- (LL) --node[circle, fill=white, draw=white]{} (UL);
\draw[double] (UL) -- ++( 135:1.0);
\draw (UR) -- ++(  45:1.0) node[anchor=south west] {};
\draw (LR) -- ++( -45:1.0) node[anchor=north west] {};
\draw (LL) -- ++(-135:1.0) node[anchor=north east] {};
\draw[red, dashed] (0,0)+(180:0.5) -- ++(180:1.0);
\drawregionvariables{$x_3$}{$\hat{x}_1^-\;$}{$\hat{x}_1$}{$\; x_2$}{$x_0$}
\end{tikzpicture}
} &= \frac{1}{\hat{x}_{01}^2 x_{02}^2 x_{03}^2 (x_{01}^+)^2} \;, \cr
\raisebox{-0.48\height}{
\begin{tikzpicture}[thick,scale=0.6]
\drawULempty
\drawURempty
\drawLRempty
\drawLLempty
\draw (UL) -- (UR) -- (LR) --node[circle, fill=white, draw=white]{} (LL) -- (UL);
\draw[double] (UL) -- ++( 135:1.0);
\draw (UR) -- ++(  45:1.0) node[anchor=south west] {};
\draw (LR) -- ++( -45:1.0) node[anchor=north west] {};
\draw (LL) -- ++(-135:1.0) node[anchor=north east] {};
\draw[red, dashed] (0,0)+(-90:0.5) -- ++(-90:1.0);
\drawregionvariables{\raisebox{-0.4cm}{$\hat{x}_1^-$}}{$x_2^-\;$}{$x_2$}{$\; x_3$}{$x_0$}
\end{tikzpicture}
} &= \frac{1}{x_{02}^2 x_{03}^2 (x_{01}^+)^2 (x_{02}^+)^2} \;, &
\raisebox{-0.48\height}{
\begin{tikzpicture}[thick,scale=0.6]
\drawULempty
\drawURempty
\drawLRempty
\drawLLempty
\draw (UL) -- (UR) --node[circle, fill=white, draw=white]{} (LR) -- (LL) -- (UL);
\draw[double] (UL) -- ++( 135:1.0);
\draw (UR) -- ++(  45:1.0) node[anchor=south west] {};
\draw (LR) -- ++( -45:1.0) node[anchor=north west] {};
\draw (LL) -- ++(-135:1.0) node[anchor=north east] {};
\draw[red, dashed] (0,0)+(  0:0.5) -- ++(  0:1.0);
\drawregionvariables{\raisebox{-0.4cm}{$x_2^-$}}{$x_3^-\;$}{$x_3$}{$\; \hat{x}_1^-$}{$x_0$}
\end{tikzpicture}
} &= \frac{1}{x_{03}^2 (x_{01}^+)^2 (x_{02}^+)^2 (x_{03}^+)^2} \;, \cr
\raisebox{-.45\height}{
\begin{tikzpicture}[thick, scale=0.75]
\coordinate (T1) at ( 0.5 ,  0.87);
\coordinate (T2) at ( 0.5 , -0.87);
\coordinate (T3) at (-1.0 ,  0.0 );
\draw (T1) -- (T2) -- (T3) --node[circle, fill=white, draw=white]{} (T1);
\draw[red, dashed] (0,0)+(120:0.3) -- ++(120:0.7);
\node[] (Tlbl3) at (-0.6,  0.8) {$\hat{x}_1$};
\node[] (Tlbl1) at ( 1.2,  0.0) {$x_3$};
\node[] (Tlbl2) at (-0.6, -0.8) {$\hat{x}_1^-$};
\node[] (Tlbl0) at ( 0.0,  0.0) {$x_0$};
\draw (T1) -- ++( 20:0.6) node[anchor=west] {};
\draw (T1) -- ++(-20:0.6) node[anchor=west] {};
\draw (T2) -- ++(0:0.6) node[anchor=west] {};
\draw[double] (T3) -- ++(180:0.6) node[anchor=east] {};
\end{tikzpicture}
} &= \frac{1}{x_{01}^2 x_{03}^2 (\hat{x}_{01}^+)^2} \;, & 
\raisebox{-.45\height}{
\begin{tikzpicture}[thick, scale=0.75]
\coordinate (T1) at ( 0.5 ,  0.87);
\coordinate (T2) at ( 0.5 , -0.87);
\coordinate (T3) at (-1.0 ,  0.0 );
\draw (T1) -- (T2) --node[circle, fill=white, draw=white]{} (T3) -- (T1);
\draw[red, dashed] (0,0)+(240:0.3) -- ++(240:0.7);
\node[] (Tlbl3) at (-0.6,  0.8) {$\hat{x}_1$};
\node[] (Tlbl1) at ( 1.2,  0.0) {$x_3$};
\node[] (Tlbl2) at (-0.6, -0.8) {$\hat{x}_1^-$};
\node[] (Tlbl0) at ( 0.0,  0.0) {$x_0$};
\draw (T1) -- ++( 20:0.6) node[anchor=west] {};
\draw (T1) -- ++(-20:0.6) node[anchor=west] {};
\draw (T2) -- ++(0:0.6) node[anchor=west] {};
\draw[double] (T3) -- ++(180:0.6) node[anchor=east] {};
\end{tikzpicture}
} &= \frac{1}{\hat{x}_{01}^2 x_{03}^2 (x_{01}^+)^2} \;, \cr
\raisebox{-.45\height}{
\begin{tikzpicture}[thick, scale=0.75]
\coordinate (T1) at ( 0.5 ,  0.87);
\coordinate (T2) at ( 0.5 , -0.87);
\coordinate (T3) at (-1.0 ,  0.0 );
\draw (T1) -- (T2) -- (T3) --node[circle, fill=white, draw=white]{} (T1);
\draw[red, dashed] (0,0)+(120:0.3) -- ++(120:0.7);
\node[] (Tlbl3) at (-0.6,  0.8) {$\hat{x}_1$};
\node[] (Tlbl1) at ( 1.2,  0.0) {$x_2$};
\node[] (Tlbl2) at (-0.6, -0.8) {$\hat{x}_1^-$};
\node[] (Tlbl0) at ( 0.0,  0.0) {$x_0$};
\draw (T1) -- ++(0:0.6) node[anchor=west] {};
\draw (T2) -- ++( 20:0.6) node[anchor=west] {};
\draw (T2) -- ++(-20:0.6) node[anchor=west] {};
\draw[double] (T3) -- ++(180:0.6) node[anchor=east] {};
\end{tikzpicture}
} &= \frac{1}{x_{01}^2 x_{02}^2 (\hat{x}_{01}^+)^2} \;, & 
\raisebox{-.45\height}{
\begin{tikzpicture}[thick, scale=0.75]
\coordinate (T1) at ( 0.5 ,  0.87);
\coordinate (T2) at ( 0.5 , -0.87);
\coordinate (T3) at (-1.0 ,  0.0 );
\draw (T1) -- (T2) --node[circle, fill=white, draw=white]{} (T3) -- (T1);
\draw[red, dashed] (0,0)+(240:0.3) -- ++(240:0.7);
\node[] (Tlbl3) at (-0.6,  0.8) {$\hat{x}_1$};
\node[] (Tlbl1) at ( 1.2,  0.0) {$x_2$};
\node[] (Tlbl2) at (-0.6, -0.8) {$\hat{x}_1^-$};
\node[] (Tlbl0) at ( 0.0,  0.0) {$x_0$};
\draw (T1) -- ++(0:0.6) node[anchor=west] {};
\draw (T2) -- ++( 20:0.6) node[anchor=west] {};
\draw (T2) -- ++(-20:0.6) node[anchor=west] {};
\draw[double] (T3) -- ++(180:0.6) node[anchor=east] {};
\end{tikzpicture}
} &= \frac{1}{\hat{x}_{01}^2 x_{02}^2 (x_{01}^+)^2} \;, \cr
\raisebox{-.45\height}{
\begin{tikzpicture}[thick, scale=0.75]
\coordinate (T1) at ( 0.5 ,  0.87);
\coordinate (T2) at ( 0.5 , -0.87);
\coordinate (T3) at (-1.0 ,  0.0 );
\draw (T1) --node[circle, fill=white, draw=white]{} (T2) -- (T3) -- (T1);
\draw[red, dashed] (0,0)+(  0:0.3) -- ++(  0:0.7);
\node[] (Tlbl3) at (-0.6,  0.8) {$x_3$};
\node[] (Tlbl1) at ( 1.2,  0.0) {$\hat{x}_1^-$};
\node[] (Tlbl2) at (-0.6, -0.8) {$x_3^-$};
\node[] (Tlbl0) at ( 0.0,  0.0) {$x_0$};
\draw (T1) -- ++(0:0.6) node[anchor=west] {};
\draw (T2) -- ++( 20:0.6) node[anchor=west] {};
\draw (T2) -- ++(-20:0.6) node[anchor=west] {};
\draw[double] (T3) -- ++(180:0.6) node[anchor=east] {};
\end{tikzpicture}
} &= \frac{1}{x_{03}^2  (x_{01}^+)^2(x_{03}^+)^2} \;, &
\raisebox{-.45\height}{
\begin{tikzpicture}[thick, scale=0.75]
\coordinate (T1) at ( 0.5 ,  0.87);
\coordinate (T2) at ( 0.5 , -0.87);
\coordinate (T3) at (-1.0 ,  0.0 );
\draw (T1) --node[circle, fill=white, draw=white]{} (T2) -- (T3) -- (T1);
\draw[red, dashed] (0,0)+(  0:0.3) -- ++(  0:0.7);
\node[] (Tlbl3) at (-0.6,  0.8) {$x_2$};
\node[] (Tlbl1) at ( 1.2,  0.0) {$\hat{x}_1^-$};
\node[] (Tlbl2) at (-0.6, -0.8) {$x_2^-$};
\node[] (Tlbl0) at ( 0.0,  0.0) {$x_0$};
\draw (T1) -- ++( 20:0.6) node[anchor=west] {};
\draw (T1) -- ++(-20:0.6) node[anchor=west] {};
\draw (T2) -- ++(  0:0.6) node[anchor=west] {};
\draw[double] (T3) -- ++(180:0.6) node[anchor=east] {};
\end{tikzpicture}
} &= \frac{1}{x_{02}^2  (x_{01}^+)^2(x_{02}^+)^2} \;.
\end{align}
As done in the previous section, we shift $x_0$ appropriately on each term to collect an overall $1/x_{01}^2$ factor. The sum of all the residues reads
\begin{align}
\mathcal{I}_{\mathrm{cut}} = -\frac{F_{3,0}^{(0)}}{2x_{01}^2} &\left[ \frac{(\hat{x}_{21}^+)^2 \hat{x}_{13}^2}{x_{02}^2 x_{03}^2 (\hat{x}_{01}^+)^2} + \frac{(\hat{x}_{21}^+)^2 \hat{x}_{13}^2}{(x_{03}^-)^2 (x_{02}^-)^2 (\hat{x}_{01}^-)^2} + \frac{(\hat{x}_{21}^+)^2(x_{32}^+)^2}{(x_{02}^-)^2 x_{02}^2 (x_{03}^-)^2} + \frac{\hat{x}_{13}^2(x_{32}^+)^2}{x_{03}^2 x_{02}^2 (x_{03}^-)^2} \right. \cr
&+ \frac{(x_{32}^+)^2+(\hat{x}_{21}^+)^2}{ (x_{03}^-)^2 x_{03}^2}+\frac{\hat{x}_{13}^2+(x_{32}^+)^2}{ x_{02}^2 (\hat{x}_{01}^+)^2} + \frac{(x_{32}^+)^2+(\hat{x}_{21}^+)^2}{ x_{03}^2(\hat{x}_{01}^+)^2} + \frac{(x_{32}^+)^2+\hat{x}_{13}^2}{ x_{02}^2 (x_{02}^-)^2} \cr
&\left.+ \frac{(\hat{x}_{21}^+)^2+(x_{32}^+)^2}{(x_{03}^-)^2 (\hat{x}_{01}^-)^2} + \frac{(x_{32}^+)^2+\hat{x}_{13}^2}{(x_{02}^-)^2 (\hat{x}_{01}^-)^2}\right] \;.
\end{align}

We will now show that the above can be obtained through the forward limit of the five-point NMHV form factor. We start from the general expression for the NMHV form factor \eqref{genNHMV} and we consider the five-point case
\begin{equation}
\tilde{F}^{(0)}_{5,1}=R'_{135}+R'_{145}+R'_{155}+R''_{135}+R'_{134}+R'_{133}+R'_{144}+R''_{131}+R''_{141} \;.
\end{equation}
We then consider the forward limit of legs $4$ and $5$ by setting
\begin{align}\label{forward5pt}
\lambda_5 &\to -\lambda_{4}\, ,  & \tilde \lambda_5 \, , &\to \tilde \lambda_{4} & \eta_5 &= \eta_{4}  \;.
\end{align}
Analogously to the previous case, only some $R$-invariants give a non-vanishing contribution after the fermionic integration,
\begin{align}
\lim_{p_5 \rightarrow -p_4} \int\mathrm{d}^4 \eta_4  \, R'_{135} &=  \frac{\delta^{(8)}(\mathrm{q}) \, [1\,2]^2}{(p_{12}-p_4)^2 \, p_{14}^2 \, [4|q\ket{3}\braket{3\,4} } \;, \cr
\lim_{p_5 \rightarrow -p_4} \int\mathrm{d}^4 \eta_4  \, R'_{145} &= -\frac{\delta^{(8)}(\mathrm{q}) \, q^4}{(q-p_4)^2 \braket{1\,2}\braket{2\,3}[4|q\ket{3}[4|q\ket{4} \braket{1\,4}} \;, \cr
\lim_{p_5 \rightarrow -p_4} \int\mathrm{d}^4 \eta_4  \, R'_{155} &= -\frac{\delta^{(8)}(\mathrm{q}) \, q^4}{(q+p_4)^2 \braket{1\,2}\braket{2\,3}[4|q\ket{1}[4|q\ket{4} \braket{3\,4}} \;, \cr
\lim_{p_5 \rightarrow -p_4} \int\mathrm{d}^4 \eta_4  \, R''_{135} &= \frac{\delta^{(8)}(\mathrm{q}) \, [2\,3]^2}{(p_{23}+p_4)^2 \, p_{34}^2 \,[4|q\ket{1}\braket{4\,1} } \;.
\end{align}
As usual, the result of BCFW recursion relations contains spurious poles. By making use of the kinematic identities
\begin{align}
\braket{2\,4}[4|q\ket{3} [3\,2]&=s_{24}s_{23}+\tfrac12 (s_{13}s_{24}-s_{12} s_{34}+s_{14}s_{23}) \;, \cr
\braket{1\,4}[4|q\ket{3} [3\,1]&=s_{14}s_{13}+\tfrac12 (s_{13}s_{24}-s_{12} s_{34}+s_{14}s_{23}) \;, \cr
\braket{2\,4}[4|q\ket{1} [1\,2]&=s_{12}s_{24}+\tfrac12 (s_{12}s_{34}-s_{14} s_{23}+s_{13}s_{24}) \;, \cr
\braket{3\,4}[4|q\ket{1} [1\,2]&=s_{13}s_{34}+\tfrac12 (s_{12}s_{34}-s_{14} s_{23}+s_{13}s_{24}) \;,
\end{align}
and after some partial fractioning, we can write the sum of the four terms above as
\begin{align}\label{Ires}
\mathcal{I}_{\mathrm{forw}} = \;&-\frac{F^{(0)}_{3,0}}{2}\left[\frac{s_{12}s_{13} }{s_{1,2,-4} s_{1,-4} s_{34}}+\frac{s_{23} s_{12}}{ s_{1,-4} s_{1,2,-4} s_{1,2,3,-4}}+\frac{s_{23}s_{13} }{s_{2,3,4} s_{1,-4} s_{34}}+\frac{s_{23} s_{12}}{s_{34} s_{2,3,4} s_{1,2,3,4}}\right. \\
&\left.+\frac{s_{13}+s_{23}}{s_{34} s_{1,2,-4}}+\frac{s_{12}+s_{13}}{s_{1,-4} s_{1,2,3,-4}}+\frac{s_{13}+s_{23}}{s_{1,2,-4}s_{1,2,3,-4}}+\frac{s_{13}+s_{12}}{s_{1,-4} s_{2,3,4}}+\frac{s_{23}+s_{13}}{s_{34} s_{1,2,3,4}}+\frac{s_{13}+s_{12}}{s_{2,3,4}s_{1,2,3,4}}\right] \, . \nonumber
\end{align}
If we now identify $x_5=x_0^-$, $x_4=x_1^-$ and perform the shift $x_1 \mapsto \hat{x}_1$, {\it i.e.}\ if we set
\begin{align}
 s_{1,-4}&=x_{02}^2\, ,  &  s_{1,2,-4}&=x_{03}^2 \, ,& s_{1,2,3,-4}&=(\hat{x}_{01}^+)^2\, , \cr
 s_{34}&=(x_{03}^-)^2\, , & s_{2,3,4}&= (x_{02}^-)^2\, , & s_{1,2,3,4}&=(\hat{x}_{01}^-)^2 \, ,\cr
 s_{12}&=\hat{x}_{13}^2 \, ,& s_{23}&=(\hat{x}_{21}^+)^2 \, ,& s_{13}&=(x_{32}^+)^2 \ , 
\end{align}
we arrive at
\begin{align}
\mathcal{I}_{\mathrm{cut}} = \frac{1}{x_{01}^2} \, \hat{\mathcal{I}}_{\mathrm{forw}} \;.
\end{align}
The complete integrand is then obtained by including the contribution from the first line in \eqref{recursion}, where the
corresponding residue is due to the overall tree-level MHV form factor $F^{(0)}_{3,0}$ which leads to the factorisation 
depicted in \eqref{EQ:MHVb-NMHV_factorisation}.

\subsection{All-line loop recursion relation}	
In \cite{Bullimore:2010dz}, an all-line, or MHV recursion relation for one-loop amplitude integrands was formulated, as an application of the integrand loop recursion of \cite{ArkaniHamed:2010kv} combined with the tree-level MHV recursion of \cite{Risager:2005vk}. In this section we show how this MHV loop recursion is extended to include also form factors. This is based on the application of MHV rules \cite{Cachazo:2004kj} to form factors \cite{Brandhuber:2011tv}, which can be immediately extended to one-loop form factors using the formalism developed in \cite{Brandhuber:2004yw,Bedford:2004py,Bedford:2004nh,Brandhuber:2005kd}.

To formulate the all-line recursion relation we employ the all-line shift of \cite{Risager:2005vk}, where all the region momenta are deformed \cite{Bullimore:2010dz}:
\begin{align}\label{xishift}
\hat{x}_i^\bullet(z) \equiv x_i^\bullet + z \, \rho_i \, \zeta \;, 
\end{align}
where  
\begin{align}\label{qi}
\rho_{i} \equiv \frac{r_i \lambda_{i-1} - r_{i-1} \lambda_i}{\langle i-1\, i \rangle} \;, 
\end{align}
and the $r_i$s are non-vanishing complex numbers which ensure that all the region momenta receive a non-vanishing shift.
They obey the periodicity condition $r_i=r_{i \pm n}=r_i^\pm$ in order to ensure that the deformed kinematic configuration remains periodic since under this condition $\rho_{i \pm n} = \rho_i^\pm=\rho_i$. Finally $\zeta_{\dot\alpha}$ is a constant reference spinor. 
It can easily be checked that the corresponding shifts of the spinors of the particles are
\begin{align}\label{holo}
\hat{\lambda}_i &\equiv \lambda_i \;, &
\hat{\tilde\lambda}_i &\equiv \tilde\lambda_i + z \, \zeta \, \frac{r_i \langle i-2\, i+1\rangle - r_{i-1} \langle i\, i+1\rangle - r_{i+1} \langle i-1\, i\rangle}{\langle i-1\, i\rangle \langle i\, i+1\rangle} \;, 
\end{align}
confirming that these are MHV diagram-type shifts: only the anti-holomorphic spinors of the particles' momenta are shifted. As a consequence, since the MHV form factor vertices are holomorphic, the only dependence on $z$ occurs through the propagators which also receive a $z$-dependent~shift. 

To explain this concretely, we focus on the MHV diagram expansion for the simplest case, namely the one-loop two-point (or Sudakov) form factor, which contains already the main features of a generic computation. 

\begin{figure}[htb]
\begin{align*}
&\raisebox{-.47\height}{
\begin{tikzpicture}[thick, scale=0.8]
\drawblobs
\drawupperlower
\draw[double] (blobL) -- ++( 180:1) node[anchor=east] {};
\draw (blobR) -- ++(  45:1) node[anchor=south west] {$1$};
\draw (blobR) -- ++( -45:1) node[anchor=north west] {$2$};
\node[] at (-1.2, +0.8) {$x_1$};
\node[] at (2, 0) {$x_2$};
\node[] at (-1.2, -0.8) {$x_1^-$};
\node[] at (0, 0) {${x_0}$};
\end{tikzpicture}
} &
&\raisebox{-.47\height}{
\begin{tikzpicture}[thick, scale=0.8]
\drawblobs
\drawupperlower
\draw[double] (blobL) -- ++( 180:1) node[anchor=east] {};
\draw (blobR) -- ++(  45:1) node[anchor=south west] {$2$};
\draw (blobR) -- ++( -45:1) node[anchor=north west] {$1$};
\node[] at (-1.2, +0.8) {$x_2$};
\node[] at (2, 0) {$x_1^-$};
\node[] at (-1.2, -0.8) {$x_2^-$};
\node[] at (0, 0) {${x_0}$};
\end{tikzpicture}
} \cr
&\raisebox{-.47\height}{
\begin{tikzpicture}[thick, scale=0.8]
\drawblobs
\drawupperlower
\draw[double] (blobL) -- ++(-135:1) node[anchor=east] {};
\draw (blobL) -- ++( 135:1) node[anchor=south east] {$1$};
\draw (blobR) -- ++(0:1) node[anchor=west] {$2$};
\node[] at (1.2, +0.8) {$x_2$};
\node[] at (-2, 0) {$x_1$};
\node[] at (1.2, -0.8) {$x_1^-$};
\node[] at (0, 0) {${x_0}$};
\end{tikzpicture}
} &
&\raisebox{-.47\height}{
\begin{tikzpicture}[thick, scale=0.8]
\drawblobs
\drawupperlower
\draw[double] (blobL) -- ++(-135:1) node[anchor=east] {};
\draw (blobL) -- ++( 135:1) node[anchor=south east] {$2$};
\draw (blobR) -- ++(0:1) node[anchor=west] {$1$};
\node[] at (1.2, +0.8) {$x_1^-$};
\node[] at (-2, 0) {$x_2$};
\node[] at (1.2, -0.8) {$x_2^-$};
\node[] at (0, 0) {${x_0}$};
\end{tikzpicture}
}
\end{align*}
\caption{The four one-loop MHV diagrams contributing to the one-loop Sudakov form factor.}
\label{fig:one-loop-Sudakov-MHV}
\end{figure}
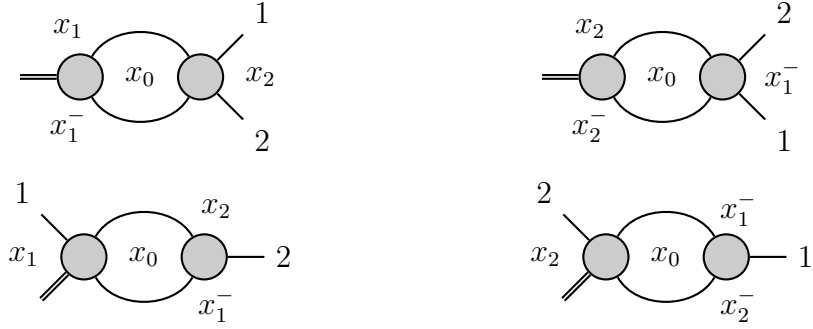

The expansion of a Sudakov form factor in terms of MHV diagrams is given in Figure \ref{fig:one-loop-Sudakov-MHV}.
Because the form factor insertion carries no colour, there are two possible types of diagrams, namely with $q=p_1+p_2$ between particles $1$ and $2$, and between particles $2$ and $1$. 
Note the appearance in the second line of that figure of diagrams that have a vanishing two-particle cut, but are nevertheless important to guarantee that the final result is independent of che choice of the reference spinor, as explicitly shown in one-loop MHV amplitude examples in \cite{Brandhuber:2004yw,Bedford:2004py,Bedford:2004nh} and later shown in full generality in \cite{Brandhuber:2005kd} using the cancellation of forward scattering singularities in supersymmetric theories. 

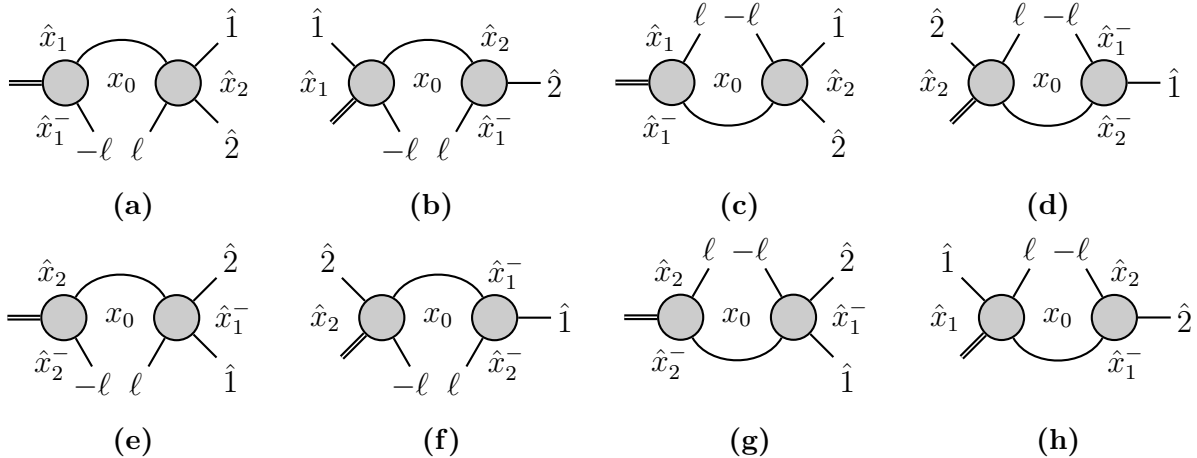
\begin{figure}[htb]
\begin{subfigure}[t]{.24\linewidth}
\centering
\begin{tikzpicture}[thick, scale=0.75]
\drawblobs
\drawupper
\drawlowerell
\draw[double] (blobL) -- ++( 180:1);
\draw (blobR) -- ++(  45:1) node[above right=-2pt] {$\hat{1}$};
\draw (blobR) -- ++( -45:1) node[below right=-2pt] {$\hat{2}$};
\node[] at (-1.2, +0.8) {$\hat{x}_1$};
\node[] at (2, 0) {$\hat{x}_2$};
\node[] at (-1.2, -0.8) {$\hat{x}_1^-$};
\node[] at (0, 0) {$x_0$};
\end{tikzpicture}
\caption{}
\end{subfigure}%
\begin{subfigure}[t]{.24\linewidth}
\centering
\begin{tikzpicture}[thick, scale=0.75]
\drawblobs
\drawupper
\drawlowerell
\draw[double] (blobL) -- ++(-135:1);
\draw (blobL) -- ++( 135:1) node[above left=-2pt] {$\hat{1}$};
\draw (blobR) -- ++(0:1) node[right=-2pt] {$\hat{2}$};
\node[] at (1.2, +0.8) {$\hat{x}_2$};
\node[] at (-2, 0) {$\hat{x}_1$};
\node[] at (1.2, -0.8) {$\hat{x}_1^-$};
\node[] at (0, 0) {$x_0$};
\end{tikzpicture}
\caption{}
\end{subfigure}
\begin{subfigure}[t]{.24\linewidth}
\centering
\raisebox{.03\height}{
\begin{tikzpicture}[thick, scale=0.75]
\drawblobs
\drawupperell
\drawlower
\draw[double] (blobL) -- ++( 180:1);
\draw (blobR) -- ++(  45:1) node[above right=-2pt] {$\hat{1}$};
\draw (blobR) -- ++( -45:1) node[below right=-2pt] {$\hat{2}$};
\node[] at (-1.2, +0.8) {$\hat{x}_1$};
\node[] at (2, 0) {$\hat{x}_2$};
\node[] at (-1.2, -0.8) {$\hat{x}_1^-$};
\node[] at (0, 0) {$x_0$};
\end{tikzpicture}
}
\caption{}
\end{subfigure}
\begin{subfigure}[t]{.24\linewidth}
\centering
\raisebox{.11\height}{
\begin{tikzpicture}[thick, scale=0.75]
\drawblobs
\drawupperell
\drawlower
\draw[double] (blobL) -- ++(-135:1);
\draw (blobL) -- ++( 135:1) node[above left=-2pt] {$\hat{2}$};
\draw (blobR) -- ++(0:1) node[right=-2pt] {$\hat{1}$};
\node[] at (1.2, +0.8) {$\hat{x}_1^-$};
\node[] at (-2, 0) {$\hat{x}_2$};
\node[] at (1.2, -0.8) {$\hat{x}_2^-$};
\node[] at (0, 0) {$x_0$};
\end{tikzpicture}
}
\caption{}
\end{subfigure}
\\
\begin{subfigure}[t]{.24\linewidth}
\centering
\begin{tikzpicture}[thick, scale=0.75]
\drawblobs
\drawupper
\drawlowerell
\draw[double] (blobL) -- ++( 180:1);
\draw (blobR) -- ++(  45:1) node[above right=-2pt] {$\hat{2}$};
\draw (blobR) -- ++( -45:1) node[below right=-2pt] {$\hat{1}$};
\node[] at (-1.2, +0.8) {$\hat{x}_2$};
\node[] at (2, 0) {$\hat{x}_1^-$};
\node[] at (-1.2, -0.8) {$\hat{x}_2^-$};
\node[] at (0, 0) {$x_0$};
\end{tikzpicture}
\caption{}
\end{subfigure}
\begin{subfigure}[t]{.24\linewidth}
\centering
\begin{tikzpicture}[thick, scale=0.75]
\drawblobs
\drawupper
\drawlowerell
\draw[double] (blobL) -- ++(-135:1);
\draw (blobL) -- ++( 135:1) node[above left=-2pt] {$\hat{2}$};
\draw (blobR) -- ++(0:1) node[right=-2pt] {$\hat{1}$};
\node[] at (1.2, +0.8) {$\hat{x}_1^-$};
\node[] at (-2, 0) {$\hat{x}_2$};
\node[] at (1.2, -0.8) {$\hat{x}_2^-$};
\node[] at (0, 0) {$x_0$};
\end{tikzpicture}
\caption{}
\end{subfigure}
\begin{subfigure}[t]{.24\linewidth}
\centering
\raisebox{.03\height}{
\begin{tikzpicture}[thick, scale=0.75]
\drawblobs
\drawupperell
\drawlower
\draw[double] (blobL) -- ++( 180:1);
\draw (blobR) -- ++(  45:1) node[above right=-2pt] {$\hat{2}$};
\draw (blobR) -- ++( -45:1) node[below right=-2pt] {$\hat{1}$};
\node[] at (-1.2, +0.8) {$\hat{x}_2$};
\node[] at (2, 0) {$\hat{x}_1^-$};
\node[] at (-1.2, -0.8) {$\hat{x}_2^-$};
\node[] at (0, 0) {$x_0$};
\end{tikzpicture}
}
\caption{}
\end{subfigure}
\begin{subfigure}[t]{.24\linewidth}
\centering
\raisebox{.11\height}{
\begin{tikzpicture}[thick, scale=0.75]
\drawblobs
\drawupperell
\drawlower
\draw[double] (blobL) -- ++(-135:1);
\draw (blobL) -- ++( 135:1) node[above left=-2pt] {$\hat{1}$};
\draw (blobR) -- ++(0:1) node[right=-2pt] {$\hat{2}$};
\node[] at (1.2, +0.8) {$\hat{x}_2$};
\node[] at (-2, 0) {$\hat{x}_1$};
\node[] at (1.2, -0.8) {$\hat{x}_1^-$};
\node[] at (0, 0) {$x_0$};
\end{tikzpicture}
}
\caption{}
\end{subfigure}
\caption{Recombination of single-cut  diagrams contributing to the one-loop recursion. The first line contributes to $F(1,2,\ell, -\ell)$ while the second to $F(2,1,\ell, -\ell)$. Diagrams \textbf{(a,b,d)} and \textbf{(c)} are accompanied by a propagator $1/(x_{01}^+)^2$ and $1/x_{01}^2$, respectively; while for diagrams \textbf{(e,f)}, and \textbf{(g,h)}, the corresponding propagators are $1/(x_{02}^+)^2$ and $1/x_{02}^2$, respectively.}
\label{fig:one-loop-Sudakov-MHV-single-cut-bis-open-up-again}
\end{figure}

Consider now a generic MHV diagram and perform a shift of the region momenta as in \eqref{xishift}. Because the MHV vertices are holomorphic, the shift \eqref{holo} does not affect them. Hence, the only $z$-dependence occurs through the shifted propagators in a generic MHV diagram. In the case at hand there will be two shifted propagators, and the residue theorem takes the form 
\begin{align}
\label{zero}
\int \frac{\mathrm{d}z}{z} \; {1\over (x_0 - \hat{x}_i)^2} \, {1\over (x_0 - \hat{x}_j)^2} = 0 \;, 
\end{align}
where $x_0 - \hat{x}_i$ and $x_0 - \hat{x}_j$ are the shifted momenta in the propagators belonging to the same MHV diagram, with the shifts given by \eqref{xishift}. For instance, in the first MHV diagram in Figure \ref{fig:one-loop-Sudakov-MHV}, these would be $x_0 - \hat{x}_1$ and $x_0 - \hat{x}_1^-$.
Furthermore, note that
\begin{align}
(x_0-\hat{x}_i)^2 = - \langle \rho_i | x_0-x_i | \zeta] \, (z-z_i)
\ , 
\end{align}
with 
\begin{align}\label{zi}
z_i = \frac{(x_0-x_i)^2}{\langle \rho_i | x_0-x_i | \zeta]} \ ,
\end{align}
ensuring that there is no pole at infinity -- and hence the validity of \eqref{zero}.
\begin{figure}[htb]
\begin{align*}
&\;\;\,\raisebox{-.47\height}{
\begin{tikzpicture}[thick, scale=0.75]
\drawblobs
\drawuppercross
\drawlowercut
\draw[double] (blobL) -- ++( 180:1);
\draw (blobR) -- ++(  45:1) node[above right=-2pt] {$\hat{1}$};
\draw (blobR) -- ++( -45:1) node[below right=-2pt] {$\hat{2}$};
\node[] at (-1.2, +0.8) {$\hat{x}_1$};
\node[] at (2, 0) {$\hat{x}_2$};
\node[] at (-1.2, -0.8) {$\hat{x}_1^-$};
\node[] at (0, 0) {$x_0$};
\end{tikzpicture}
} &
&\kern 0.2em\raisebox{-.47\height}{
\begin{tikzpicture}[thick, scale=0.75]
\drawblobs
\drawuppercut
\drawlowercross
\draw[double] (blobL) -- ++( 180:1);
\draw (blobR) -- ++(  45:1) node[above right=-2pt] {$\hat{1}$};
\draw (blobR) -- ++( -45:1) node[below right=-2pt] {$\hat{2}$};
\node[] at (-1.2, +0.8) {$\hat{x}_1$};
\node[] at (2, 0) {$\hat{x}_2$};
\node[] at (-1.2, -0.8) {$\hat{x}_1^-$};
\node[] at (0, 0) {$x_0$};
\end{tikzpicture}
} &
&\kern 0.2em\raisebox{-.47\height}{
\begin{tikzpicture}[thick, scale=0.75]
\drawblobs
\drawuppercross
\drawlowercut
\draw[double] (blobL) -- ++( 180:1);
\draw (blobR) -- ++(  45:1) node[above right=-2pt] {$\hat{2}$};
\draw (blobR) -- ++( -45:1) node[below right=-2pt] {$\hat{1}$};
\node[] at (-1.2, +0.8) {$\hat{x}_2$};
\node[] at (2, 0) {$\hat{x}_1^-$};
\node[] at (-1.2, -0.8) {$\hat{x}_2^-$};
\node[] at (0, 0) {$x_0$};
\end{tikzpicture}
} &
&\kern 0.2em\raisebox{-.47\height}{
\begin{tikzpicture}[thick, scale=0.75]
\drawblobs
\drawuppercut
\drawlowercross
\draw[double] (blobL) -- ++( 180:1);
\draw (blobR) -- ++(  45:1) node[above right=-2pt] {$\hat{2}$};
\draw (blobR) -- ++( -45:1) node[below right=-2pt] {$\hat{1}$};
\node[] at (-1.2, +0.8) {$\hat{x}_2$};
\node[] at (2, 0) {$\hat{x}_1^-$};
\node[] at (-1.2, -0.8) {$\hat{x}_2^-$};
\node[] at (0, 0) {$x_0$};
\end{tikzpicture}
} \cr
&\raisebox{-.47\height}{
\begin{tikzpicture}[thick, scale=0.75]
\drawblobs
\drawuppercross
\drawlowercut
\draw[double] (blobL) -- ++(-135:1);
\draw (blobL) -- ++( 135:1) node[above left=-2pt] {$\hat{1}$};
\draw (blobR) -- ++(0:1) node[right=-2pt] {$\hat{2}$};
\node[] at (1.2, +0.8) {$\hat{x}_2$};
\node[] at (-2, 0) {$\hat{x}_1$};
\node[] at (1.2, -0.8) {$\hat{x}_1^-$};
\node[] at (0, 0) {$x_0$};
\end{tikzpicture}
} &
&\!\!\!\raisebox{-.47\height}{
\begin{tikzpicture}[thick, scale=0.75]
\drawblobs
\drawuppercut
\drawlowercross
\draw[double] (blobL) -- ++(-135:1);
\draw (blobL) -- ++( 135:1) node[above left=-2pt] {$\hat{1}$};
\draw (blobR) -- ++(0:1) node[right=-2pt] {$\hat{2}$};
\node[] at (1.2, +0.8) {$\hat{x}_2$};
\node[] at (-2, 0) {$\hat{x}_1$};
\node[] at (1.2, -0.8) {$\hat{x}_1^-$};
\node[] at (0, 0) {$x_0$};
\end{tikzpicture}
} &
&\!\!\!\raisebox{-.47\height}{
\begin{tikzpicture}[thick, scale=0.75]
\drawblobs
\drawuppercross
\drawlowercut
\draw[double] (blobL) -- ++(-135:1);
\draw (blobL) -- ++( 135:1) node[above left=-2pt] {$\hat{2}$};
\draw (blobR) -- ++(0:1) node[right=-2pt] {$\hat{1}$};
\node[] at (1.2, +0.8) {$\hat{x}_1^-$};
\node[] at (-2, 0) {$\hat{x}_2$};
\node[] at (1.2, -0.8) {$\hat{x}_2^-$};
\node[] at (0, 0) {$x_0$};
\end{tikzpicture}
} &
&\!\!\!\raisebox{-.47\height}{
\begin{tikzpicture}[thick, scale=0.75]
\drawblobs
\drawuppercut
\drawlowercross
\draw[double] (blobL) -- ++(-135:1);
\draw (blobL) -- ++( 135:1) node[above left=-2pt] {$\hat{2}$};
\draw (blobR) -- ++(0:1) node[right=-2pt] {$\hat{1}$};
\node[] at (1.2, +0.8) {$\hat{x}_1^-$};
\node[] at (-2, 0) {$\hat{x}_2$};
\node[] at (1.2, -0.8) {$\hat{x}_2^-$};
\node[] at (0, 0) {$x_0$};
\end{tikzpicture}
}
\end{align*}
\caption{The single-cut diagrams contributing to the one-loop recursion. A red dotted line indicates a cut propagator, and a cross implies that the corresponding shifted propagator is evaluated on the solution to the cut of the other shifted propagator.}
\label{fig:one-loop-Sudakov-MHV-single-cut-bis}
\end{figure}
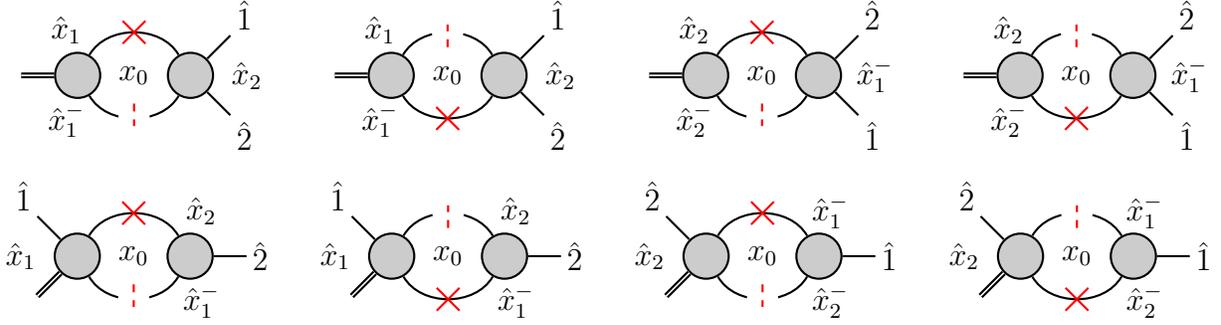
The statement of the recursion relation in this particular one-loop example is therefore nothing but 
\begin{align}\label{residues}
\frac{1}{z_i z_j} + \frac{1}{z_i (z_i-z_j)} + \frac{1}{z_j (z_j-z_i)} = 0 \;.
\end{align}
Next we  use that  
\begin{align}
\left. (x_0 - \hat{x}_i)^2 \right|_{z_j} &= - \langle \rho_i | x_0 - x_i | \zeta] (z_j - z_i)
\, , \\
\left. (x_0 - \hat{x}_j)^2 \right|_{z_i} &= - \langle \rho_j | x_0 - x_j | \zeta] (z_i - z_j) \;, 
\end{align}
as well as the standard BCFW relation 
\begin{align}\label{standardBCFW}
z_l \, \langle \rho_l | x_0 - x_l | \zeta] = (x_0 - x_l)^2
\ , 
\end{align}
where the right-hand side of \eqref{standardBCFW} is the usual pole denominator in the BCFW recursion relation. This allows us  rewrite \eqref{residues} in a transparent way:
\begin{align}\label{residues2} 
\frac{1}{(x_0 - x_i)^2 \, (x_0 - x_j)^2} = \frac{1}{(x_0 - x_i)^2 \left.(x_0 - \hat{x}_j)^2\right|_{z_i} } + \frac{1}{(x_0 - x_j)^2 \left.(x_0 - \hat{x}_i)^2\right|_{z_j}} \;. 
\end{align}
The left-hand side is nothing but a pair of unshifted scalar propagators; they are present in all one-loop MHV diagrams we are considering in Figure \ref{fig:one-loop-Sudakov-MHV}.
The first term on the right-hand side of \eqref{residues2} is evaluated on the solution $z_i$  to 
$(x_0 - \hat{x}_i)^2=0$, while the second for $(x_0 - \hat{x}_j)^2=0$. 
The effect of such terms is as for the MHV amplitude recursion  \cite{Bullimore:2010dz}, which we quickly summarise here. For the sake of concreteness, we focus on the first MHV diagram in Figure \ref{fig:one-loop-Sudakov-MHV}. Applying \eqref{residues2}, this diagram is mapped on to  two terms, namely the first two diagrams in Figure \ref{fig:one-loop-Sudakov-MHV-single-cut-bis}. The first diagram appears with a factor of 
\begin{align}
\left.{1\over (x_0 - \hat{x}_1)^2}\right|_{(x_0 - \hat{x}_1^-)^2=0} \, {1\over (x_0 - x_1^-)^2}
\ , 
\end{align}
while the second with
\begin{align}
\left. \frac{1}{(x_0 - \hat{x}_1^-)^2}\right|_{(x_0 - \hat{x}_1)^2=0} \, \frac{1}{(x_0 - x_1)^2} \;.
\end{align}
In both cases, the first monomial is the crossed propagator appearing in  the corresponding diagram in Figure~\ref{fig:one-loop-Sudakov-MHV-single-cut-bis}, evaluated on the solution to the condition that puts on shell the other propagator originally present in the one-loop MHV diagram (and decorated with a cut). The second is a multiplicative factor that will be present in the final form for the recursion relation; it has the meaning of $1/L^2$ where $L$ is the off-shell loop integration variable (more on this later). Importantly, in the first diagram the condition $(x_0 - \hat{x}_1^-)^2= (\hat{x}_{01}^+)^2=0$ puts on shell the other shifted loop momentum, opening up the propagator, and adding two particles in a forward scattering  configuration. The massless momenta of the two additional particles are $\ell$ and $-\ell$, where the on-shell momentum $\ell$ is precisely 
\begin{align}
\ell = \left. \hat{x}_1^- - x_0 \right|_{z_1^-} \;.
\end{align}
In the second diagram the roles of the two propagators are swapped, and
\begin{align}
\ell = \left. \hat{x}_1 - x_0 \right|_{z_1} \;.
\end{align}
This has the effect of performing  a single cut of the four one-loop MHV diagram of Figure~\ref{fig:one-loop-Sudakov-MHV}, which are mapped into the eight cut diagrams shown in Figure \ref{fig:one-loop-Sudakov-MHV-single-cut-bis-open-up-again}.

For convenience, we have shown the same diagrams again in Figure \ref{fig:one-loop-Sudakov-MHV-single-cut-bis-open-up-again} (but slightly reordered), and  it is clear that these diagrams split into two sets: \textbf{(a)}--\textbf{(d)} correspond to MHV diagrams contributing to the four-point NMHV form factor with particle ordering $(1,2,\ell, -\ell)$, while \textbf{(e)}--\textbf{(h)} to MHV diagrams contributing to the four-point NMHV form factor with particle ordering $(2,1,\ell, -\ell)$.  

There are two important points to discuss next -- the first one is the familiar absence of some diagrams (in the tree-level recombination into a NMHV form factor), while the second is new and characteristic of form factors. We discuss them in turn. 

{\bf 1.}~First, we focus on the first line of Figure \ref{fig:one-loop-Sudakov-MHV-single-cut-bis-open-up-again} and make the observation that summing these four diagrams one would obtain the NMHV form factor with legs $(1,2,\ell, -\ell)$,  minus some  {\it ``missing" MHV diagrams}, \textit{i.e.}\ those where particles $\ell$ and $-\ell$ belong to the same MHV vertex (with an implicit sum over all particles in the theory that can propagate along the cut leg). This class of diagrams is obviously never produced when cutting open a one-loop MHV diagram.
 
Such missing diagrams, which we have already encountered in the BCFW recursion of Section~\ref{SEC:BCFW}, have appeared in several instances \cite{CaronHuot:2010zt, ArkaniHamed:2010kv}; to the best of our knowledge, their first appearance is in Section 3.1 of \cite{Brandhuber:2005kd}, where it was shown that such diagrams vanish in this forward-scattering configuration upon performing the \mbox{(super-)sum} over the internal species, which is equivalent to performing the integration over the Grassmann variables corresponding to the internal legs $\int \mathrm{d}^4\eta_\ell$.
As a consequence, the four diagrams in the first line of Figure \ref{fig:one-loop-Sudakov-MHV-single-cut-bis-open-up-again} reconstruct by themselves  the four-point tree-level NMHV form factor $F^{(0)}_{\rm NMHV}(1,2,\ell,-\ell)$. 

{\bf 2.}~The second important point, which we have also encountered already in Section~\ref{SEC:BCFW}, is a specific feature arising for form factors. Indeed, as recalled earlier, the single-cut diagrams in Figure \ref{fig:one-loop-Sudakov-MHV-single-cut-bis-open-up-again} are accompanied by particular denominators of the form $1/L^2$ (with $L$ being an off-shell loop momentum) as demanded by \eqref{residues2}. For the eight diagrams in that figure these are 
\begin{align}\label{se}
\frac{1}{(x_{01}^+)^2} \;, \quad \frac{1}{(x_{01}^+)^2} \;, \quad \frac{1}{x_{01}^2} \;, \quad \frac{1}{(x_{01}^+)^2} \;, 
\end{align}
for the first four diagrams (first line), and 
\begin{align}
\frac{1}{(x_{02}^+)^2} \;, \quad \frac{1}{(x_{02}^+)^2} \;, \quad \frac{1}{x_{02}^2} \;, \quad \frac{1}{(x_{02}^+)^2} \;, 
\end{align}
for the remaining four diagrams (second line). In the recursion for amplitudes, due to the planarity of the diagrams there is no such ambiguity and only one denominator appears. 

Correspondingly, in this form factor recursion the meaning of $\ell$ is different in these diagrams. Indeed we have, for the first four diagrams 
\begin{align}
\ell = \hat{x}_{{1}0}^- \;, \quad \ell =
 \hat{x}_{{1}0}^- \;, \quad \ell = \hat{x}_{{1}0} \;, \quad \ell = \hat{x}_{{1}0}^-
\;,
\end{align}
and 
\begin{align}
\ell = \hat{x}_{{2}0}^- \;, \quad \ell = \hat{x}_{{2}0}^- \;, \quad \ell = \hat{x}_{{2}0} \;, \quad \ell = \hat{x}_{{2}0}
\;, 
\end{align}
for the last ones. 
Crucially, these two issues are fixed by allowing shifts in the integration variable $x_0$ by $q$, which we can do in diagram \textbf{(c)}, as well as in diagram \textbf{(g)}, \textbf{(h)}.
More in detail, we can focus on diagrams \textbf{(a)}, \textbf{(b)}, \textbf{(d)} in Figure \ref{fig:one-loop-Sudakov-MHV-single-cut-bis-open-up-again}. By performing the change of integration variable $x_0 \to x_0 -q$ with $q=x_1-x_1^-$, the new denominator becomes $(x_0 -q-x_1^-)^2=(x_0-x_1)^2$, that is as in diagram \textbf{(c)}. Furthermore, the meaning  of $\ell$ becomes the same as in the diagram \textbf{(c)}. 
Indeed, before the shift we have: 
\begin{align}
\text{diagrams}\, \textbf{(a)}, \textbf{(b)}, \textbf{(d)}: \quad &\ell=\left. (\hat{x}_3-x_0)\right|_{z_1^-} \;, & \text{diagram}\, \textbf{(c)}: 
\quad &\ell=\left. (\hat{x}_1-x_0)\right|_{z_1}\ . 
\end{align}
After the shift in $x_0$ we get, using \eqref{xishift},  
\begin{align}\label{shh}
\left. (\hat{x}_1^--x_0)\right|_{z_1^-} \to x_1 -x_0 +  \rho_1^- \zeta \left.z_1^- \right|_{x_0 \to x_0 - q}  \;.
\end{align}
We can now show that the right-hand side of \eqref{shh} is nothing but $\left.(\hat{x}_1-x_0)\right|_{z_1}$. First, we note that $\rho_i= \rho_i^-$, from the definition \eqref{qi} and the fact that $\lambda_{i+n} = \lambda_i$. Hence we simply have to show that 
\begin{align}
\left.z_1^- \right|_{x_0 \to x_0 - q} = z_1 \;, 
\end{align}
which follows from \eqref{zi} and $\rho_1^- = \rho_1$.

In conclusion, after performing appropriate shifts in the diagrams \textbf{(a)}, \textbf{(b)}, \textbf{(d)} we bring all diagrams in the first line of Figure \ref{fig:one-loop-Sudakov-MHV-single-cut-bis-open-up-again} to have $\ell = \left.\hat{x}_{{1}0}\right|_{z_1}$, while for the second line of the same figure we can perform appropriate shift  to arrive at $\ell = \left.\hat{x}_{{2}0}\right|_{z_2}$. 
In conclusion, the recursion relation here has the form 
\begin{align}\label{finalSud}
F^{(1)}_{2,0} (1,2) = \int \mathrm{d}^d x_0 \; \mathrm{d}^4 \eta_\ell \left[ \frac{F^{(0)}_{4,1} (\hat{1}, \hat{2}, \ell, -\ell)}{(x_0-x_1)^2} + \frac{F^{(0)}_{4,1}(\hat{1}, \ell, -\ell, \hat{2})}{(x_0-x_2)^2} \right] \;,
\end{align}
where the first term is evaluated on the solution to $\hat{x}_{0{1}}^2=0$ while the second for $\hat{x}_{0{2}}^2=0$. In the first term $\ell=\left.\hat{x}_{{1}0}\right|_{z_1}$, while in the second $\ell=\left.\hat{x}_{{2}0}\right|_{z_2}$. A few final remarks are in order. 

{\bf 1.}~First, we would like to   comment on the arbitrariness of the assignments of region momenta. 
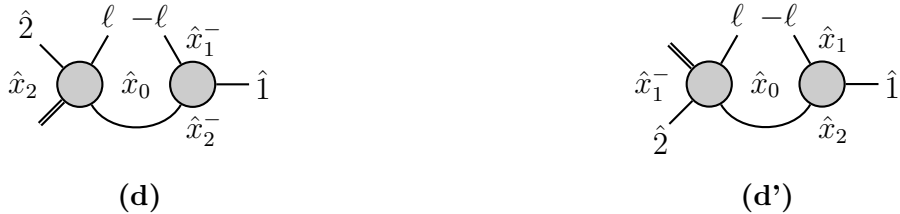
\begin{figure}[htb]
\begin{subfigure}[b]{.5\linewidth}
\centering
\raisebox{0.045\height}{
\begin{tikzpicture}[thick, scale=0.75]
\drawblobs
\drawupperell
\drawlower
\draw[double] (blobL) -- ++(-135:1);
\draw (blobL) -- ++( 135:1) node[above left=-2pt] {$\hat{2}$};
\draw (blobR) -- ++(0:1) node[right=-2pt] {$\hat{1}$};
\node[] at (1.2, +0.8) {$\hat{x}_1^-$};
\node[] at (-2, 0) {$\hat{x}_2$};
\node[] at (1.2, -0.8) {$\hat{x}_2^-$};
\node[] at (0, 0) {$\hat{x}_0$};
\end{tikzpicture}
}
\caption*{\textbf{(d)}}
\end{subfigure}
\begin{subfigure}[b]{.5\linewidth}
\centering
\begin{tikzpicture}[thick, scale=0.75]
\drawblobs
\drawupperell
\drawlower
\draw[double] (blobL) -- ++( 135:1);
\draw (blobL) -- ++(-135:1) node[below left=-4pt] {$\hat{2}$};
\draw (blobR) -- ++(0:1) node[right=-2pt] {$\hat{1}$};
\node[] at (1.2, +0.8) {$\hat{x}_1$};
\node[] at (-2, 0) {$\hat{x}_1^-$};
\node[] at (1.2, -0.8) {$\hat{x}_2$};
\node[] at (0, 0) {$\hat{x}_0$};
\end{tikzpicture}
\caption*{\textbf{(d')}}
\end{subfigure}
\caption{Two different (but equivalent) ways to depict diagram \textbf{(d)} of the previous Figure. In the first depiction $\ell=\hat{x}_{10}^-$ while in the second $\ell=\hat{x}_{10}$, which differs from the previous one by $q$.}
\label{arbitrariness}
\end{figure}
In order to making it manifest, we have drawn diagram \textbf{(d)} of Figure \ref{fig:one-loop-Sudakov-MHV-single-cut-bis-open-up-again} in two different ways in Figure \ref{arbitrariness}. The form factor insertion is colour blind, hence there is no reason to prefer one to the other. The assignments of region momenta are modified correspondingly. 
Using \textbf{(d')} instead of \textbf{(d)} would change \eqref{se} into a more ``symmetric" 
\begin{align}\label{se2}
\frac{1}{(x_{01}^+)^2} \;, \quad \frac{1}{(x_{01}^+)^2} \;, \quad \frac{1}{x_{01}^2} \;, \quad \frac{1}{x_{01}^2} \;. 
\end{align}
The point to make is that there is no preferred choice -- both  give the same answer for the integrand thanks to the possibility of shifting $x_0$ by $q$. Moreover, we could write various terms in the recursion \eqref{finalSud} using variables belonging to different periods; as an example, the first term on the right-hand side of \eqref{finalSud} could have been written as 
\begin{align}
\frac{F^{(0)}_{4,0} (\hat{1}, \hat{2}, \ell, -\ell)}{(x_0-x_1^-)^2}\ , 
\end{align}
with $\ell$ now being given by $\ell= \left.(\hat{x}_1^--x_0)\right|_{z_1^-}$, where we also recall that $x_1^- = x_1-q$. 

{\bf 2.}~We also note that the one-loop recursion relation for the two-point case \eqref{finalSud}  extends immediately to an arbitrary number of  points. For the one-loop MHV form factors the recursion has the form 
\begin{align}
F^{(1)}_{n,0} (1, \ldots , n) = \int \mathrm{d}^d x_0 \; \mathrm{d}^4 \eta_{\ell} \ 
\Big[ 
\frac{F^{(0)}_{n+2,1} (\hat{1}, \hat{2}, \ldots , \hat{n}, \ell, -\ell)}{(x_0-x_1)^2} +\cdots +  \frac{F^{(0)}_{n+2,1} (\hat{n}, \hat{1}, \ldots , \widehat{n-1}, \ell, -\ell)}{(x_0-x_n)^2}
\Big] \;.
\end{align}

{\bf 3.}~In the case of $\mathrm{N^kMHV}$ form factors with $k \geq 1$, also the familiar terms corresponding to standard factorisation appear: the recursion then reads
\begin{align}
\begin{split}
F^{(1)}_{n,k} (1, \ldots , n) &= \int \mathrm{d}^d x_0 \; \mathrm{d}^4 \eta_{\ell} \ 
\Big[ 
\sum_{i=1}^n \frac{F^{(0)}_{n+2,k+1} (\hat{1}, \hat{2}, \ldots , \widehat{i-1}, \ell, -\ell, \hat{i}, \ldots , \hat{n})}{(x_0-x_i)^2} +
\cr
&+\sum_{l,i,k_{\mathrm{L}}}  \int \mathrm{d}^4\eta_{\ell} \; \bigg[{F}^{(l)}_{i,k_{\mathrm{L}}}(\hat{x}_1,\ldots,\hat{x}_i) \; \frac{1}{(x_{i1}^+)^2} \; {A}^{(1-l)}_{n-i+2,k_{\mathrm{R}}}(\hat{x}_1,\hat{x}_i,\dots,\hat{x}_n) 
\cr
&\kern6.1em + {A}^{(l)}_{i,k_{\mathrm{L}}}(\hat{x}_1,\ldots,\hat{x}_i) \; \frac{1}{(x_{i1})^2} \; {F}^{(1-l)}_{n-i+2,k_{\mathrm{R}}}(\hat{x}_1,\hat{x}_i,\dots,\hat{x}_n) \Big] \;,
%
\end{split}
\end{align}
where, as in \eqref{recursion}, $l=0,1$, $i=2,\ldots,n-1$ and $k_{\mathrm{L}}+k_{\mathrm{R}}=k-1$ with $k_{\mathrm{L}}, k_{\mathrm{R}} \ge 0$.

{\bf 4.}~Bonus relation: In \cite{ArkaniHamed:2008gz}, it was noted that thanks to their $1/z^2$ fall-off at infinity, $\mathcal{N}=8$ supergravity amplitudes at tree level satisfy a bonus recursion relation of the type 
\begin{align}
\int \mathrm{d}z \; \mathcal{A}(z) = 0 \;,
\end{align}
where $ \mathcal{A}(z)$ is the shifted $\mathcal{N}=8$ superamplitude with a supersymmetric two-line shift. Here we make the rather simple  observation that because all internal propagators in a one-loop MHV diagram receive a shift (and therefore fall off as $1/z$ for large $z$), we will therefore have bonus recursion relations. With at least two propagators, a generic one-loop integrand will behave, under the all-line shift, as $1/z^2$. Again focusing on the one-loop Sudakov form factor, in this case the bonus  relation reads
\begin{align}
\label{Bogus}
0 = \int \mathrm{d}^d x_0 \; \bigg[ 
\frac{F^{(0)}_{4,1} (\hat{1}, \hat{2}, \ell, -\ell)}{D_1} + \frac{F^{(0)}_{4,1} ( \hat{1}, \ell, -\ell, \hat{2})}{D_2} \bigg] \;, 
\end{align}
where $D_i \equiv (x_0 - x_i)^2/z_i = \langle \rho_i | (x_0 - x_i) | \zeta] $, where we used \eqref{zi}.  
Note that this is no longer a recursion relation for one-loop integrands, rather a constraint on the NMHV form factors at tree level.

\section*{Acknowledgements}
We would like to thank \"{O}mer Gurdo\u{g}an, Florian Loebbert, Jan Plefka and Alexander Tumanov for interesting discussions. 
The work of LB is supported by a Marie Sk\l{}odowska-Curie Individual Fellowship under grant agreement No.~749909.
LB would like to thank the ``Dipartimento di Fisica'' of Torino University for kind hospitality during the initial phase of this project.
The work of AB and GT was supported by the Science and Technology Facilities Council (STFC) Consolidated Grant ST/L000415/1  
\textit{``String theory, gauge theory \& duality"}. This project has received funding from the European Union's Horizon 2020 research and innovation programme under the Marie Sk\l{}odowska-Curie grant agreement 
No.~764850 {\it ``SAGEX"}.
GT is grateful to the Alexander von Humboldt Foundation for support through a Friedrich Wilhelm Bessel Research Award, and to the Institute for Physics and IRIS Adlershof at Humboldt University, Berlin, for their warm hospitality.

\newpage
\appendix
\section{Conventions and notation} \label{conventions}
The fundamental building blocks used in this paper are the three-point superamplitudes and the two-point, or Sudakov form factor: 
\begin{align}
A_{3,0}^{(0)} &=
\raisebox{-.44\height}{
\begin{tikzpicture}[thick]
  \blackdot (MHVb) at (0.0, 0.0) {};
  \draw (MHVb) -- ++( 120:0.8) node[anchor=east] {$1$};
  \draw (MHVb) -- ++(   0:0.8) node[anchor=west] {$2$};
  \draw (MHVb) -- ++(-120:0.8) node[anchor=east] {$3$};
\end{tikzpicture}
} = \mathrm{i} \, \frac{\delta^{(8)}(\lambda_1 \eta_1 + \lambda_2 \eta_2 + \lambda_3 \eta_3)}{\langle1\,2\rangle \langle2\,3\rangle \langle3\,1\rangle} \, , \cr
A_{3,1}^{(0)} &=
\raisebox{-.44\height}{
\begin{tikzpicture}[thick]
  \whitedot (MHVb) at (0.0, 0.0) {};
  \draw (MHVb) -- ++( 120:0.8) node[anchor=east] {$1$};
  \draw (MHVb) -- ++(   0:0.8) node[anchor=west] {$2$};
  \draw (MHVb) -- ++(-120:0.8) node[anchor=east] {$3$};
\end{tikzpicture}
} = -\mathrm{i} \, \frac{\delta^{(4)}([2\,3]\eta_1 + [3\,1]\eta_2 + [1\,2]\eta_3)}{[1\,2][2\,3][3\,1]}\, ,  \cr
F_{2,0}^{(0)} &=
\raisebox{-.44\height}{
\begin{tikzpicture}[thick]
  \coordinate (FF) at (0.0, 0.0);
  \draw (FF) -- ++(  60:0.8) node[anchor=west] {$1$};
  \draw (FF) -- ++( -60:0.8) node[anchor=west] {$2$};
  \draw[double] (FF) -- ++( 180:0.8) node[anchor=east] {};
\end{tikzpicture}
} = \frac{\delta^{(8)}(\mathrm{q})}{\langle1\,2\rangle \langle2\,1\rangle}\, .
\end{align}
The off-shell leg of the form factor, which is indicated by a double line, carries incoming momentum $q$ and supermomentum $\gamma$, with
\begin{align}
q = \sum_{i=1}^n p_i \;, \qquad \mathrm{q} = \sum_{i=1}^n \mathrm{q}_i - \gamma \;.
\end{align}
Note that $\mathcal{F}_2^{\text{MHV}}$ is the minimal supersymmetric form factor of the chiral half of the protected stress-tensor multiplet (for details see \cite{Brandhuber:2011tv}) and $\gamma$ labels different components of this multiplet.

Because there is a notion of ordering for on-shell legs, the kinematics of a $n$-point form factor can be realised in terms of dual coordinates by specifying a set of $x^{\alpha\dot{\alpha}}_i$ and $\theta^{A \alpha}_i$ such that
\begin{align}
x^{\alpha\dot{\alpha}}_i - x^{\alpha\dot{\alpha}}_{i+1} &= p_i^{\alpha\dot{\alpha}} = \lambda_i^\alpha\widetilde{\lambda}_i^{\dot{\alpha}} \;, \\
\theta^{A \alpha}_i - \theta^{A \alpha}_{i+1} &= \mathrm{q}_i^{A \alpha} = \eta_i^A \lambda_i^{\alpha} \; .
\end{align}
More generally one has, for $i<j$,
\begin{align}
p_i + p_{i+1} + \cdots + p_j = x_i - x_{j+1} \equiv x_{i\,j+1} \;,
\end{align}
and similarly for the $\theta_i$ variables. If $q\neq0$ the dual coordinates will not describe a closed polygon. Cyclicity can be fully realised by introducing periodic images for the points $x_i$ with
\begin{align}
x^{[m]}_{i} &= x_i + mq \;, &
\theta^{[m]}_{i} &= \theta_i + m\gamma \;,
\end{align}
with $m\in\mathbb{Z}$. This generates a periodic segmented line in the space of dual coordinates. For the particular case $m=\pm 1$ we use the notation
\begin{align}
x^\pm_i &= x_i\pm q \;, & \theta^\pm_i &= \theta_i \pm \gamma \;.
\end{align}

The same kinematic configuration can be encoded in terms of momentum-twistor variables 
\cite{Hodges:2009hk}
since edges of the periodic line are light rays in dual space. The incidence relation
\begin{align}
\mu^{\dot{\alpha}}_i = x^{\alpha\dot{\alpha}}_i \lambda_{i\,\alpha} = x^{\alpha\dot{\alpha}}_{i+1} \lambda_{i\,\alpha}
\end{align}
fixes the components of the twistor $Z_i = (\lambda_i, \mu_i)$, and the ambiguity in the choice of the spinor-helicity variables $(\lambda_i, \widetilde{\lambda}_i)$ now translates to the fact that $Z_i$ are interpreted as  projective coordinates in twistor space $\mathbb{T}\simeq\mathbb{CP}^3$.
Periodicity is implemented \cite{Brandhuber:2011tv} by the condition
\begin{align}
\lambda_{i+n \, a} = \lambda_{i \, \alpha} \;, \qquad \mu^{\dot{\alpha}}_{i+n} = \mu^{\dot{\alpha}}_i - q^{\alpha\dot{\alpha}} \lambda_{i \, \alpha} \;.
\end{align}
This can be seen as the finite translation generated by
\begin{align}
\mathsf{P}_{\alpha\dot{\alpha}} = \lambda_\alpha \frac{\partial}{\partial \mu^{\dot{\alpha}}} \;.
\end{align}

\section{Details on the tree-level NMHV form factor}\label{detailsNMHV}
In this appendix we outline the computation of NMHV tree-level form factors using on-shell diagrams. We use a BCFW shift of the $[1 \, 2\rangle$ kind. For an $n$-point form factors the recursion gives
\begin{align}\label{EQ:BCFW_NMHV}
F^{(0)}_{n,1} = \;&\sum_{i=4}^{n}
\raisebox{-.475\height}{
\begin{tikzpicture}[thick, scale=0.8]
  \drawLLwhite
  \drawULblack
  \drawUR{$\scriptstyle{0}$}
  \drawLR{$\scriptstyle{0}$}
  \drawboxinternallines
  \draw (UL) -- ++( 135:0.8) node[above left=-2pt] {$2$}; 
  \draw[double] (UR) -- ++(   0:0.8);
  \draw (UR) -- ++(  90:0.8) node[above=-2pt] {$3$};
  \draw (UR) -- ++(  45:0.8) node[above right=-2pt] {$i-1$};
  \draw[dotted] (UR)+(  80:0.6) to [bend left=45] ++( 50:0.6);
  \draw (LR) -- ++(   0:0.8) node[right=-2pt] {$i$};
  \draw[dotted] (LR)+( -10:0.6) to [bend left=45] ++( -80:0.6);
  \draw (LR) -- ++( -90:0.8) node[below=-2pt] {$n$}; 
  \draw (LL) -- ++(-135:0.8) node[below left=-2pt] {$1$};
\end{tikzpicture}
}
+ \sum_{i=5}^{n}
\raisebox{-.46\height}{
\begin{tikzpicture}[thick, scale=0.8]
  \drawLLwhite
  \drawULblack
  \drawUR{$\scriptstyle{0}$}
  \drawLR{$\scriptstyle{0}$}
  \drawboxinternallines
  \draw (UL) -- ++( 135:0.8) node[above left=-2pt] {$2$}; 
  \draw[double] (LR) -- ++( -90:0.8);
  \draw (UR) -- ++(  90:0.8) node[above=-2pt] {$3$};
  \draw (UR) -- ++(   0:0.8) node[right=-2pt] {$i-1$};
  \draw[dotted] (UR)+(  80:0.6) to [bend left=45] ++( 10:0.6);
  \draw (LR) -- ++(   0:0.8) node[right=-2pt] {$i$};
  \draw[dotted] (LR)+( -10:0.6) to [bend left=45] ++( -40:0.6);
  \draw (LR) -- ++( -45:0.8) node[below right=-2pt] {$n$}; 
  \draw (LL) -- ++(-135:0.8) node[below left=-2pt] {$1$};
\end{tikzpicture}
} \cr
& +
\raisebox{-.46\height}{
\begin{tikzpicture}[thick, scale=0.8]
  \drawLLwhite
  \drawULblack
  \drawUR{$\scriptstyle{0}$}
  \drawLRempty
  \drawboxinternallines
  \draw (UL) -- ++( 135:0.8) node[above left=-2pt] {$2$}; 
  \draw[double] (LR) -- ++( -45:0.8);
  \draw (UR) -- ++(  90:0.8) node[above=-2pt] {$3$};
  \draw (UR) -- ++(   0:0.8) node[right=-2pt] {$n$};
  \draw[dotted] (UR)+(  80:0.6) to [bend left=45] ++( 10:0.6);
  \draw (LL) -- ++(-135:0.8) node[below left=-2pt] {$1$};
\end{tikzpicture}
} +
\raisebox{-.505\height}{
\begin{tikzpicture}[thick, scale=0.8]
  \drawLLwhite
  \drawULblack
  \drawURempty
  \drawLR{$\scriptstyle{0}$}
  \drawboxinternallines
  \draw (UL) -- ++( 135:0.8) node[above left=-2pt] {$2$}; 
  \draw[double] (UR) -- ++(  45:0.8);
  \draw (LR) -- ++(   0:0.8) node[right=-2pt] {$3$};
  \draw[dotted] (LR)+( -10:0.6) to [bend left=45] ++( -80:0.6);
  \draw (LR) -- ++( -90:0.8) node[below=-2pt] {$n$}; 
  \draw (LL) -- ++(-135:0.8) node[below left=-2pt] {$1$};
\end{tikzpicture}
} +
\raisebox{-.49\height}{
\begin{tikzpicture}[thick, scale=0.8]
  \drawLLwhite
  \drawULblack
  \drawURwhite
  \drawLR{$\scriptstyle{1}$}
  \drawboxinternallines
  \draw (UL) -- ++( 135:0.8) node[above left=-2pt] {$2$}; 
  \draw[double] (LR) -- ++( -90:0.8);
  \draw (UR) -- ++(  45:0.8) node[above right=-2pt] {$3$};
  \draw (LR) -- ++(   0:0.8) node[right=-2pt] {$4$};
  \draw[dotted] (LR)+( -10:0.6) to [bend left=45] ++( -40:0.6);
  \draw (LR) -- ++( -45:0.8) node[below right=-2pt] {$n$}; 
  \draw (LL) -- ++(-135:0.8) node[below left=-2pt] {$1$};
\end{tikzpicture}
}
\end{align}
Here the recursion is represented in terms of so-called \emph{BCFW bridges}. The last diagram can be written in terms of $R$-invariants by recursively inserting the NMHV $(n-1)$-point form factor in the lower-right corner. 

To understand how many $R$-invariants contribute to that diagram,  one can use the following argument. An $n$-point NMHV form factor is expressed in terms of $2n-5$ diagrams containing products of MHV amplitudes and form factors and one diagram containing the combination of a NMHV $(n-1)$-point form factor and a $\overline{\text{MHV}}$ three-point amplitude. If one denotes with $a_n$ the number of $R$-invariants associated to the $n$-point NMHV form factor, one can replace the NMHV $(n-1)$-point form factor with its $a_{n-1}$ $R$-invariants. This  gives a recursive relation, 
\begin{align}
a_n = a_{n-1} + 2n-5 \;,
\end{align}
which is solved by
\begin{align}
a_n = (n-2)^2 \;. 
\end{align}
Consequently, the diagram involving a NMHV form factor times a $\overline{\text{MHV}}$ three-point amplitude should decompose into $(n-3)^2$ box coefficients. The precise combination for a $[1 \, 2\rangle$ shift is 
\begin{equation}
F^{(0)}_{n,1} = F^{(0)}_{n,0}\left(\sum_{j=3}^{n} \sum_{i=3}^{j} R'_{1ij}+\sum_{j=5}^{n+1}\sum_{i=3}^{j-2} R''_{1ij}\right) \;,
\end{equation}
where we make the identification $n+1\!\sim\!1$.
The number of $R$-invariants in this expression~is
\begin{equation}
\underbrace{\frac{(n-2)(n-1)}{2}}_{R'}+\underbrace{\frac{(n-2)(n-3)}{2}}_{R''}=(n-2)^2 \;.
\end{equation}

Finally we want to illustrate how the NMHV$\times\overline{\text{MHV}}$ diagram can be written in terms of the \mbox{$R$-invariants} introduced in this paper. An elegant way to achieve this are on-shell diagrams. We show how this works for the four-point and five-point form factor. Together with the usual rules for on-shell diagrammatics that are used for amplitudes, namely
\begin{align}
\raisebox{-.45\height}{
\begin{tikzpicture}[thick, scale=0.7]
  \whitedot (LL) at (-\boxsize, -\boxsize) {};
  \blackdot (UL) at (-\boxsize, +\boxsize) {};
  \whitedot (UR) at (+\boxsize, +\boxsize) {};
  \blackdot (LR) at (+\boxsize, -\boxsize) {};
  \drawboxinternallines
  \draw (UL) -- ++( 135:0.8); 
  \draw (UR) -- ++(  45:0.8);
  \draw (LR) -- ++( -45:0.8);
  \draw (LL) -- ++(-135:0.8);
\end{tikzpicture}
} &=
\raisebox{-.45\height}{
\begin{tikzpicture}[thick, scale=0.7]
  \blackdot (LL) at (-\boxsize, -\boxsize) {};
  \whitedot (UL) at (-\boxsize, +\boxsize) {};
  \blackdot (UR) at (+\boxsize, +\boxsize) {};
  \whitedot (LR) at (+\boxsize, -\boxsize) {};
  \drawboxinternallines
  \draw (UL) -- ++( 135:0.8); 
  \draw (UR) -- ++(  45:0.8);
  \draw (LR) -- ++( -45:0.8);
  \draw (LL) -- ++(-135:0.8);
\end{tikzpicture}
} \;,
&
\raisebox{-.38\height}{
\begin{tikzpicture}[thick, scale=0.7]
  \blackdot (L) at (-\boxsize, 0) {};
  \blackdot (R) at (+\boxsize, 0) {};
  \draw (L) -- ++( 135:0.8);
  \draw (L) -- ++(-135:0.8);
  \draw (R) -- ++(  45:0.8);
  \draw (R) -- ++( -45:0.8);
  \draw (L) -- (R);
\end{tikzpicture}
} &=
\raisebox{-.45\height}{
\begin{tikzpicture}[thick, scale=0.7]
  \blackdot (D) at (0,-\boxsize) {};
  \blackdot (U) at (0,+\boxsize) {};
  \draw (U) -- ++( 135:0.8);
  \draw (U) -- ++(  45:0.8);
  \draw (D) -- ++( -45:0.8);
  \draw (D) -- ++(-135:0.8);
  \draw (D) -- (U);
\end{tikzpicture}
} \;,
&
\raisebox{-.38\height}{
\begin{tikzpicture}[thick, scale=0.7]
  \whitedot (L) at (-\boxsize, 0) {};
  \whitedot (R) at (+\boxsize, 0) {};
  \draw (L) -- ++( 135:0.8);
  \draw (L) -- ++(-135:0.8);
  \draw (R) -- ++(  45:0.8);
  \draw (R) -- ++( -45:0.8);
  \draw (L) -- (R);
\end{tikzpicture}
} &=
\raisebox{-.45\height}{
\begin{tikzpicture}[thick, scale=0.7]
  \whitedot (D) at (0,-\boxsize) {};
  \whitedot (U) at (0,+\boxsize) {};
  \draw (U) -- ++( 135:0.8);
  \draw (U) -- ++(  45:0.8);
  \draw (D) -- ++( -45:0.8);
  \draw (D) -- ++(-135:0.8);
  \draw (D) -- (U);
\end{tikzpicture}
} \;,
\end{align}
we use the fundamental vertices associated with the off-shell leg insertion,
\begin{align}
\raisebox{-.45\height}{
\begin{tikzpicture}[thick, scale=1.0]
\node[circle, fill=black!20, draw, minimum size=\blobsize, inner sep=2pt] (B) at ( 0.0, 0.0) {$\scriptstyle{0}$};
\draw[double] (B) -- ++(180:0.87);
\draw (B) -- ++( 45:0.87) node[above right=-2pt] {$1$};
\draw (B) -- ++(  0:0.87) node[right=-2pt] {$2$};
\draw (B) -- ++(-45:0.87) node[below right=-2pt] {$3$};
\end{tikzpicture}
} &=
\raisebox{-.46\height}{
\begin{tikzpicture}[thick, scale=1.0]
\coordinate (L) at (-0.5, 0.0);
\whitedot (U) at ( 0.0, 0.5) {};
\blackdot (R) at ( 0.5, 0.0) {};
\whitedot (D) at ( 0.0,-0.5) {};
\draw (U) -- (R) -- (D) -- (L) -- (U);
\draw[double] (L) -- ++(180:0.5);
\draw (U) -- ++( 45:0.5) node[right=-2pt] {$1$};
\draw (R) -- ++(  0:0.5) node[right=-2pt] {$2$};
\draw (D) -- ++(-45:0.5) node[right=-2pt] {$3$};
\end{tikzpicture}
} \;, &
\raisebox{-.45\height}{
\begin{tikzpicture}[thick, scale=1.0]
\node[circle, fill=black!20, draw, minimum size=\blobsize, inner sep=2pt] (B) at ( 0.0, 0.0) {$\scriptstyle{1}$};
\draw[double] (B) -- ++(180:0.87);
\draw (B) -- ++( 45:0.87) node[above right=-2pt] {$1$};
\draw (B) -- ++(  0:0.87) node[right=-2pt] {$2$};
\draw (B) -- ++(-45:0.87) node[below right=-2pt] {$3$};
\end{tikzpicture}
} &=
\raisebox{-.46\height}{
\begin{tikzpicture}[thick, scale=1.0]
\coordinate (L) at (-0.5, 0.0);
\blackdot (U) at ( 0.0, 0.5) {};
\whitedot (R) at ( 0.5, 0.0) {};
\blackdot (D) at ( 0.0,-0.5) {};
\draw (U) -- (R) -- (D) -- (L) -- (U);
\draw[double] (L) -- ++(180:0.5);
\draw (U) -- ++( 45:0.5) node[right=-2pt] {$1$};
\draw (R) -- ++(  0:0.5) node[right=-2pt] {$2$};
\draw (D) -- ++(-45:0.5) node[right=-2pt] {$3$};
\end{tikzpicture}
} \;.
\end{align}

If we take, for example, $F^{(0)}_{4,1}$, with the above one can easily show that
\begin{align}\label{EQ:NMHV_onshell}
\raisebox{-.49\height}{
\begin{tikzpicture}[thick, scale=0.8]
  \drawLLwhite
  \drawULblack
  \drawURwhite
  \drawLR{$\scriptstyle{1}$}
  \drawboxinternallines
  \draw (UL) -- ++( 135:0.8) node[above left=-2pt] {$2$}; 
  \draw[double] (LR) -- ++( -90:0.8);
  \draw (UR) -- ++(  45:0.8) node[above right=-2pt] {$3$};
  \draw (LR) -- ++(   0:0.8) node[right=-2pt] {$4$}; 
  \draw (LL) -- ++(-135:0.8) node[below left=-2pt] {$1$};
\end{tikzpicture}
} =
\raisebox{-.47\height}{
\begin{tikzpicture}[thick, scale=0.9]
\whitedot (LL) at (-1.0, -0.5) {};
\blackdot (LM) at (-0.0, -0.5) {};
\whitedot (LR) at (+1.0, -0.5) {};
\blackdot (UL) at (-1.0, +0.5) {};
\whitedot (UM) at (+0.0, +0.5) {};
\blackdot (UR) at (+1.0, +0.5) {};
\coordinate (OS) at (+0.5,  1.0);
\draw (UM) --(UL) -- (LL) -- (LM) -- (UM) --  (OS) -- (UR) -- (LR) -- (LM);
\draw (UL) -- ++( 135:0.7) node[above left=-2pt] {$3$};
\draw (LL) -- ++(-135:0.7) node[below left=-2pt] {$2$};
\draw (LR) -- ++( -45:0.7) node[below right=-2pt] {$1$};
\draw (UR) -- ++(  45:0.7) node[above right=-2pt] {$4$};
\draw[double] (OS) -- ++(  90:0.6) {};
\end{tikzpicture}
} =
\raisebox{-.46\height}{
\begin{tikzpicture}[thick, scale=0.8]
  \drawLLwhite
  \drawLRblack
  \drawURempty
  \drawUL{$\scriptstyle{0}$}
  \drawboxinternallines
  \draw (UL) -- ++(  90:0.8) node[above=-2pt] {$3$};
  \draw (UL) -- ++( 180:0.8) node[left=-2pt] {$2$};
  \draw[double] (UR) -- ++(  45:0.8);
  \draw (LR) -- ++( -45:0.8) node[below right=-2pt] {$4$};
  \draw (LL) -- ++(-135:0.8) node[below left=-2pt] {$1$};
\end{tikzpicture}
} \;,
\end{align}
which allows us to identify the last term in the recursion as an $R$-invariant and explicitly check \eqref{genNHMV}. Similarly, for the $n=5$ case the last term in \eqref{EQ:BCFW_NMHV} can be recast as
\begin{align}
\raisebox{-.46\height}{
\begin{tikzpicture}[thick, scale=0.8]
  \drawLLwhite
  \drawUL{$\scriptstyle{0}$}
  \drawURempty
  \drawLR{$\scriptstyle{0}$}
  \drawboxinternallines
  \draw (UL) -- ++(  90:0.8) node[above=-2pt] {$3$};
  \draw (UL) -- ++( 180:0.8) node[left=-2pt] {$2$};
  \draw[double] (UR) -- ++(  45:0.8);
  \draw (LR) -- ++(   0:0.8) node[right=-2pt] {$4$};
  \draw (LR) -- ++( -90:0.8) node[below=-2pt] {$5$}; 
  \draw (LL) -- ++(-135:0.8) node[below left=-2pt] {$1$};
\end{tikzpicture}
} +
\raisebox{-.43\height}{
\begin{tikzpicture}[thick, scale=0.8]
  \drawLLwhite
  \drawUL{$\scriptstyle{0}$}
  \drawUR{$\scriptstyle{0}$}
  \drawLRblack
  \drawboxinternallines
  \draw (UL) -- ++(  90:0.8) node[above=-2pt] {$3$};
  \draw (UL) -- ++( 180:0.8) node[left=-2pt] {$2$};
  \draw (UR) -- ++(  90:0.8) node[above=-2pt] {$4$};
  \draw[double] (UR) -- ++(   0:0.8);
  \draw (LR) -- ++( -45:0.8) node[below right=-2pt] {$5$}; 
  \draw (LL) -- ++(-135:0.8) node[below left=-2pt] {$1$};
\end{tikzpicture}
} +
\raisebox{-.43\height}{
\begin{tikzpicture}[thick, scale=0.8]
  \drawLLwhite
  \drawUL{$\scriptstyle{0}$}
  \drawURempty
  \drawLRblack
  \drawboxinternallines
  \draw (UL) -- ++(  90:0.8) node[above=-2pt] {$4$};
  \draw (UL) -- ++( 135:0.8) node[above left=-2pt] {$3$};
  \draw (UL) -- ++( 180:0.8) node[left=-2pt] {$2$};
  \draw[double] (UR) -- ++(  45:0.8);
  \draw (LR) -- ++( -45:0.8) node[below right=-2pt] {$5$}; 
  \draw (LL) -- ++(-135:0.8) node[below left=-2pt] {$1$};
\end{tikzpicture}
} +
\raisebox{-.43\height}{
\begin{tikzpicture}[thick, scale=0.8]
  \drawLLwhite
  \drawUL{$\scriptstyle{0}$}
  \drawUR{$\scriptstyle{0}$}
  \drawLRempty
  \drawboxinternallines
  \draw (UL) -- ++(  90:0.8) node[above=-2pt] {$3$};
  \draw (UL) -- ++( 180:0.8) node[left=-2pt] {$2$};
  \draw (UR) -- ++(  90:0.8) node[above=-2pt] {$4$};
  \draw (UR) -- ++(   0:0.8) node[right=-2pt] {$5$};
  \draw[double] (LR) -- ++( -45:0.8); 
  \draw (LL) -- ++(-135:0.8) node[below left=-2pt] {$1$};
\end{tikzpicture}
} \;.
\end{align}

\pagebreak
\bibliographystyle{utphys}
\bibliography{remainder}

\end{document}